\documentclass[reprint,twocolumn,showpacs,superscriptaddress,footinbib,prl,floatfix,aps]{revtex4-1}
\usepackage{dcolumn,pstricks,amssymb,bm}
\usepackage[intlimits]{amsmath}
\usepackage{graphicx}
\newcommand{\tsups}[1]{\textsuperscript{#1}}  
\newcommand{\erfc}[0]{\mathrm{erfc}}
\newcommand{\erf}[0]{\mathrm{erf}}
\newcommand{\cyltran}[0]{{\scshape Cyltran}}
\newcommand{\penelope}[0]{{\scshape Penelope}}
\newcommand{\zerotozero}[0]{\mbox{$0^+\!\!\rightarrow 0^+$}}
\newcommand{\usd}[0]{\mbox{\texttt{USD}}}
\newcommand{\usda}[0]{\mbox{\texttt{USD-A}}}
\newcommand{\usdb}[0]{\mbox{\texttt{USD-B}}}

\begin{document}
%
\title{\boldmath The $\beta$ Decay of \tsups{32}Cl: Precision $\gamma$-Ray 
  Spectroscopy and a Measurement of Isospin-Symmetry Breaking}
\author{D.~Melconian}
\email[Email: ]{dmelconian@physics.tamu.edu}
\affiliation{Department of Physics, Texas A\&M University, 
  College Station, Texas 77843-4242, USA}
\affiliation{Cyclotron Institute, Texas A\&M University, 
  College Station, Texas 77843-3366, USA}
\affiliation{Department of Physics, University of Washington, Seattle, 
  Washington 98195-1560, USA}
\author{S.~Triambak}
\affiliation{Department of Physics, University of Washington, Seattle, 
  Washington 98195-1560, USA}
\affiliation{Department of Physics \& Astrophysics, 
  University of Delhi, Delhi 110 007, India}
\author{C.~Bordeanu}
\altaffiliation{Present Address: Institute of Nuclear Research of the 
  Hungarian Academy of Sciences, Debrecen, Hungary H-4001.\\
  On leave from the Horia Hulubei National Institute for Physics and 
  Nuclear Engineering, Bucharest-Magurele, Romania RO-077125.}
\affiliation{Department of Physics, University of Washington, Seattle, 
  Washington 98195-1560, USA}
\author{A.~Garc\'ia}
\affiliation{Department of Physics, University of Washington, Seattle, 
  Washington 98195-1560, USA}
\author{J.C.~Hardy}
\affiliation{Department of Physics, Texas A\&M University, 
  College Station, Texas 77843-4242, USA}
\affiliation{Cyclotron Institute, Texas A\&M University, 
  College Station, Texas 77843-3366, USA}
\author{V.E.~Iacob}
\affiliation{Cyclotron Institute, Texas A\&M University, 
  College Station, Texas 77843-3366, USA}
\author{N.~Nica}
\affiliation{Cyclotron Institute, Texas A\&M University, 
  College Station, Texas 77843-3366, USA}
\author{H.I.~Park}
\affiliation{Department of Physics, Texas A\&M University, 
  College Station, Texas 77843-4242, USA}
\affiliation{Cyclotron Institute, Texas A\&M University, 
  College Station, Texas 77843-3366, USA}
\author{G.~Tabacaru}
\affiliation{Cyclotron Institute, Texas A\&M University, 
  College Station, Texas 77843-3366, USA}
\author{L.~Trache}
\affiliation{Cyclotron Institute, Texas A\&M University, 
  College Station, Texas 77843-3366, USA}
\author{I.S.~Towner}
\affiliation{Department of Physics, Texas A\&M University, 
  College Station, Texas 77843-4242, USA}
\affiliation{Cyclotron Institute, Texas A\&M University, 
  College Station, Texas 77843-3366, USA}
\author{R.E.~Tribble}
\affiliation{Department of Physics, Texas A\&M University, 
  College Station, Texas 77843-4242, USA}
\affiliation{Cyclotron Institute, Texas A\&M University, 
  College Station, Texas 77843-3366, USA}
\author{Y.~Zhai}
\altaffiliation{Present Address: Department of Therapeutic Radiology, 
  School of Medicine, Yale University, New Haven, Connecticut, 06520, USA}
\affiliation{Department of Physics, Texas A\&M University, 
  College Station, Texas 77843-4242, USA}
\affiliation{Cyclotron Institute, Texas A\&M University, 
  College Station, Texas 77843-3366, USA}
\begin{abstract}
  \begin{description}
  \item[Background] Models to calculate small isospin-symmetry-breaking 
    effects in superallowed Fermi decays have been placed under scrutiny in 
    recent years. A stringent test of these models is to measure transitions 
    for which the correction is predicted to be large.  The decay of 
    \tsups{32}Cl decay provides such a test case.
  \item[Purpose] To improve the $\gamma$ yields following the $\beta$ decay 
    of \tsups{32}Cl and to determine the $ft$ values of the the $\beta$ 
    branches, particularly the one to the isobaric-analogue state in 
    \tsups{32}S.
  \item[Method] Reaction-produced and recoil-spectrometer-separated 
    \tsups{32}Cl is collected in tape and transported to a counting location 
    where $\beta-\gamma$ coincidences are measured with a precisely-calibrated 
    HPGe detector.
  \item[Results] The precision on the $\gamma$ yields for most of the known
    $\beta$ branches has been improved by about an order of magnitude, and many
    new transitions have been observed.  We have determined 
    \tsups{32}Cl-decay transition strengths extending up to 
    $E_x\sim11$~MeV.\ \ The $ft$ value for the decay to the isobaric-analogue 
    state in \tsups{32}S has been measured. A comparison to a shell-model 
    calculation shows good agreement.
  \item[Conclusions] We have experimentally determined the 
    isospin-symmetry-breaking correction to the superallowed transition of 
    this decay to be $(\delta_C-\delta_\mathrm{NS})_\mathrm{exp}=5.4(9)\%$, 
    significantly larger than for any other known superallowed Fermi 
    transition.  This correction agrees with a shell-model calculation, which 
    yields $\delta_C-\delta_\mathrm{NS}=4.8(5)\%$.  Our results also provide a 
    way to improve the measured $ft$ values for the $\beta$ decay of 
    \tsups{32}Ar.
  \end{description}
\end{abstract}
\date{\today}
\pacs{%
23.40.Bw, 
24.80.+y, 
29.30.Kv, 
23.20.Lv  
}
\maketitle
\section{Motivation}
The comparative half-lives of superallowed Fermi $\beta$ decays between $0^+$ 
isobaric analogue states have been the focus of intense research activity 
for many years and presently represents one of the most stringent tests of the 
Standard Model of the electroweak interaction~\cite{Hardy:09}.   The high 
precision of both experimental measurements and theoretical calculations of 
their $ft$ values set stringent limits on scalar and right-handed currents, 
verify conservation of the vector current to $\sim10^{-4}$, and determine 
the up-down element of the Cabibbo-Kobayashi-Maskawa (CKM) quark-mixing 
matrix, $V_{ud}$~\cite{HardyPRL:05,Hardy:05,Hardy:09}.  Experimentally, 
the $ft$ value of thirteen cases have been measured to $\lesssim0.3\%$; this 
places a demanding requirement on theory to attain similar precision.  
Although these transitions are intrinsically simpler to describe theoretically 
than most $\beta$ decays because they are relatively insensitive to 
nuclear-structure effects, small ($\sim1\%$) corrections must be applied 
to account for the fact that the decay occurs within the nuclear medium.  
Recently, emphasis has been placed on scrutinizing the 
nuclear-structure-dependent isospin-symmetry-breaking (ISB) corrections, 
$\delta_C$~\cite{TownerHardy:08,Auerbach:09,Liang:09,Satula:11,Miller:08,
*Miller:09}, which characterizes the degree to which the Fermi matrix 
element, $M_F$, deviates from $M_0$, its value in the limit of strict 
isospin symmetry:
\begin{align}
  |M_F|^2=|M_0|^2(1-\delta_C).\label{eq:defn_of_delta_C}
\end{align}
The 13 most precisely-measured cases mentioned previously are all isospin 
$T=1$ to $T=1$ transitions in $A=4n+2$ nuclei.  Shell-model calculations for 
these cases yield values of order $\delta_C\sim0.5\%$ 
for $A<56$~\cite{Hardy:09} and values of order $\delta_C\sim1.5\%$ for 
$A>56$~\cite{Hardy:09,hyland}.  If attention is switched to $A=4n$ nuclei, 
even larger values of $\delta_C$ are predicted, which if experimentally 
extracted, would provide an even more demanding test of such ISB 
calculations.  The reason larger values are expected in $A=4n$ nuclei is that 
the daughter analog state sits among many states of lower isospin, $T-1$.  
Some of these states have the same spin as the analog state and sizable 
isospin mixing can occur.  

In this work, we will expand on a recent Letter~\cite{melconian-32Cl-PRL} 
which discusses an extraction of the isospin-symmetry breaking correction 
in the decay of $1^+,T=1$ \tsups{32}Cl. Its Fermi decay branch feeds a 
$1^+,T\!=\!1$ state in \tsups{32}S, whose position in the spectrum at 
7001-keV excitation~\cite{triambakPRC73} is very close to a known
$1^+,T\!=\!0$ state at 7190~keV~\cite{NuclDataSheets}.\ \  As discussed 
below, a calculation of $\delta_C$ for this case yields $\delta_C=4.8(5)\%$,
a significantly larger value than those found in $A=4n+2$ nuclei.  

Another motivation for this work is related to the recently measured $T\!=\!2$ 
decay of \tsups{32}Ar~\cite{ar32-paper}.  Here calculations predict $\delta_C
=1.8\%$~\cite{SB:11}.  The ISB correction for this case extracted from the 
experimental $ft$ values was found to be $\delta_C^\mathrm{exp}=
(2.1\pm0.8)\%$~\cite{ar32-paper} and later corrected to $\delta_C^\mathrm{exp}=
(1.8\pm0.8)\%$ in Ref.~\cite{wrede}, where an improved value for the end-point 
energy was deduced.  A potentially large source of systematic uncertainty 
arising from the need to detect $\gamma$ rays in this measurement may be 
minimized using the $\gamma$ branches of \tsups{32}Cl.  This is because 
$\beta$ decay of \tsups{32}Ar is followed by the $\beta$ decay of \tsups{32}Cl 
$64.4(2)\%$ of the time~\cite{ar32-paper}.  Thus, \tsups{32}Cl provides an 
\emph{in situ} efficiency calibration which is useful to extract 
isospin-breaking information from \tsups{32}Ar.  The present work opens the 
possibility for significant improvements in the precision with which 
$\delta_C$ can be determined in the decay of \tsups{32}Ar.

\section{Experimental procedure}
The experiment was carried out at the Cyclotron Institute, Texas A\&M 
University. A primary beam of \tsups{32}S was produced by an ECR ion 
source and injected into the K500 superconducting cyclotron to accelerate 
it to $\approx$ $24.8$~MeV/nucleon.  The $\approx400$~nA \tsups{32}S beam 
exited the cyclotron and was directed towards the target chamber of the 
Momentum Achromatic Recoil Separator (MARS)~\cite{mars}. A $20$~MeV/nucleon 
secondary beam of $^{32}$Cl was produced via the inverse kinematic transfer 
reaction, $^1\mathrm{H}(^{32}\mathrm{S},n)^{32}\mathrm{Cl}$ on a LN$_2$ cooled, 
hydrogen gas target at $\approx1.4$~atm. MARS was used to spatially 
separate the reaction products, resulting in a \tsups{32}Cl beam with an 
intensity of $\sim2\times 10^{5}$~ions/s.  Beam contamination was identified 
using a position-sensitive Si-strip ($\Delta E$) detector followed by a 
silicon ($E$) detector which were placed just downstream of the MARS focal 
plane. For data collection, the Si detectors were removed and the beam exited 
the MARS beamline through a $50~\mu$m-thick Kapton window.  The beam then 
passed through a 0.3~mm thick BC404 scintillator to count the number of ions.  
Prior to being implanted into a $76~\mu$m-thick aluminized-Mylar tape which 
is part of a fast tape-transport system, the beam was passed through a set of 
Al degraders.  The thickness of the degraders was chosen to ensure that the 
activity was deposited mid-way through the tape.
The different ranges of the contaminants compared to \tsups{32}Cl 
allowed further purification; however, since we were searching for small 
branches and wanted to maximize the yields, we allowed greater contamination 
than usual of the deposited activity, accepting 91\% as our final purity of 
\tsups{32}Cl.

\begin{figure}\centering
  \includegraphics[width=0.475\textwidth]{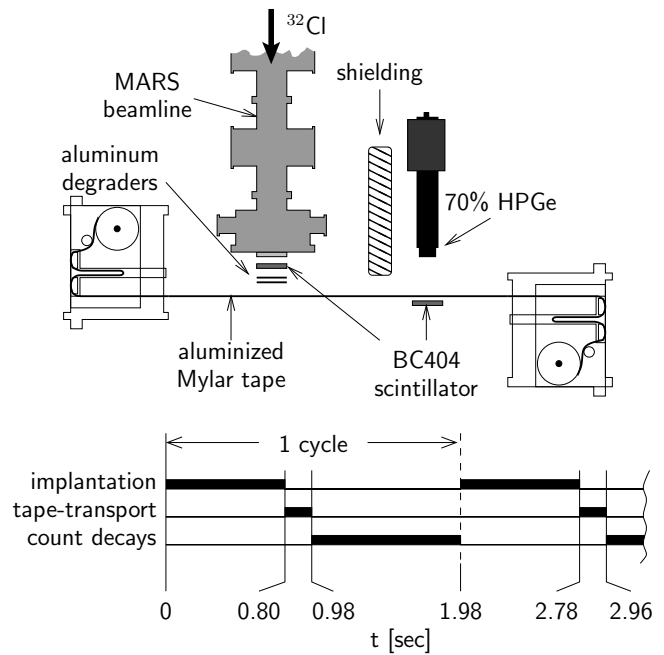}
  \caption{Schematic diagram of the end of MARS and the fast 
    tape-transport system 
    as well as a timing diagram for the experiment.  The counting time for 
    decays was generally 1~sec as indicated, but times of 2 and 4~sec were 
    also used for the diagnosis of longer-lived contaminants.}
  \label{fig1}
\end{figure}

The \tsups{32}Cl atoms were collected in an $\approx 1$~cm diameter spot on 
the tape for 0.8~s, after which the beam was interrupted and the tape-transport 
system was triggered to move the activity to a shielded counting station 
$90$~cm away. The latter was accomplished in $\approx 180$~ms. The set-up 
is shown schematically in Fig.~\ref{fig1}.  Once transported to the shielded 
area, $\beta-\gamma$ coincident data were acquired for typically 1~sec 
(83\% of the total data set).  In a few of the runs (corresponding to 11\% 
and 6\% of the data respectively) we used count times of 2~secs and 4~secs 
to check for long-lived contaminants.  The data were registered 
event-by-event by recording all $\beta-\gamma$ coincidences between a 
1-mm-thick BC404 plastic scintillator and a 70\% HPGe detector.  The 
$1.5$~inch diameter scintillator $\Delta E$ detector had a threshold of 
$40$~keV and was placed 5~mm behind the tape subtending $\approx32\%$ of the 
total solid angle for the $\beta$s. The $\gamma$ detector was placed much 
farther away: $15.1$~cm from the tape to reduce the effects of coincidence 
summing.  The $\gamma$-ray energy, the $\Delta E$ of the $\beta$, the 
coincidence time between them, and the time of the event relative to the 
beginning of the cycle were all recorded.  The typical tape cycle, outlined in 
Fig.~\ref{fig1}, was repeated continuously throughout the experiment. The 
total number of $\beta$ singles events and the total number of heavy ions (HIs) 
from MARS (detected by the first scintillator) for each cycle were determined 
from scalers and recorded.  The ratio of $\beta$ singles to HIs was used to 
veto bad cycles where, for example, the tape transport did not place the 
activity exactly in the correct location.  Of the $\approx 36\,000$ cycles 
made over the course of the experiment, 92.5\% survived the $\beta$/HI ratio 
cut.  

\subsection[gamma efficiencies]{\boldmath$\gamma$ Efficiencies}
\begin{figure}\centering
  \includegraphics[angle=90,width=0.4825\textwidth]{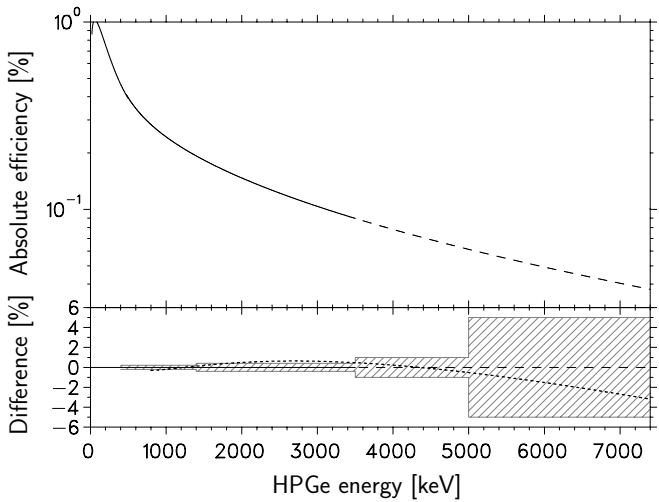}
  \caption{Top: the adopted absolute photopeak efficiency curve of the HPGe 
    used in this work.  The solid line is the measured efficiency up to 
    3.5~MeV from Refs.~\cite{hardy-AppRad,helmer-NIMA,helmer-AppRad}, while the 
    dashed line above 3.5~MeV represents an extrapolation based on \cyltran{} 
    calculations.  Bottom:  percent difference of a \penelope{} simulation 
    minus the adopted curve (dotted line) and the assigned uncertainties to 
    the efficiency curve used in the present analysis (hatched region).
    \label{fig:gamma-effs}}
\end{figure}

The extremely precise absolute photopeak efficiency calibration of the HPGe 
detector is a critical component of this experiment.  This efficiency has 
been carefully studied up to energies of 3.5~MeV as discussed in 
Refs.~\cite{hardy-AppRad,helmer-NIMA,helmer-AppRad}.  Over the range of 
$50-1400$~keV, Ref.~\cite{helmer-NIMA} discusses how the absolute efficiency 
has been calibrated to $\pm0.2\%$, and Ref.~\cite{helmer-AppRad} uses a 
combination of measurements and calculations using the 
\cyltran~\cite{cyltran} Monte Carlo code to extend the efficiency curve from 
$1.4-3.5$~MeV to $\pm0.4\%$ precision.  For this work, the highest energy 
$\gamma$ rays observed are at $7.2$~MeV, requiring us to further extend the 
photopeak efficiency curve.  Our adopted curve and its extrapolation, 
shown in the top panel of Fig.~\ref{fig:gamma-effs}, is calculated using the 
the same \cyltran{} program used in 
Refs.~\cite{hardy-AppRad,helmer-NIMA,helmer-AppRad}.  Note that above 3.5~MeV, 
the efficiency curve is no longer linear on the log-log plot as a simple 
extrapolation might suggest, instead falling off more quickly.  We additionally 
checked this extrapolation against an independent calculation based on the 
Monte Carlo radiation transport code \penelope~\cite{penelope}.  As can be 
seen in the bottom panel of Fig.~\ref{fig:gamma-effs}, there is excellent 
agreement between the two calculated photopeak efficiencies over the range of 
energies observed in this work.  As the figure also indicates, we increase the 
uncertainties in the efficiency curve, adopting conservative uncertainties of 
$\pm1$\% from 3.5--5~MeV, and $\pm5\%$ above 5~MeV.\ \ The differences between 
the \cyltran{} and \penelope{} extrapolated efficiency curves are well within 
these uncertainty ranges.

\begin{figure}\centering
  \includegraphics[angle=90,width=0.4825\textwidth]{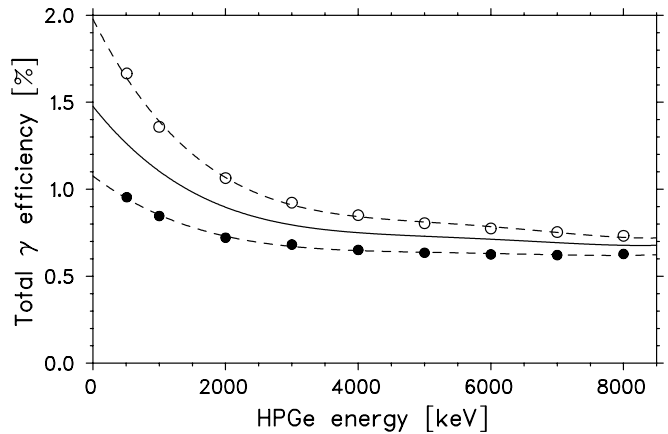}
  \caption{Total efficiency of the HPGe as simulated by 
    \penelope.  The filled circles represent the simulated points with our 
    basic geometry (detectors and Mylar tape) while the open circles 
    show the results when our basic geometry is encased in an Aluminum 
    cylinder.  The dashed lines going through each is a 4\tsups{th} order 
    polynomial fit, and the solid central line is the average of the two 
    which we take to be our total efficiency curve.  We assign an uncertainty 
    that spans the result of both simulations.  We used these curves to 
    determine systematic uncertainties associated with summing in our 
    detectors.\label{fig:tot-efficiency}}
\end{figure}

We have investigated the effects of summing in the HPGe detector using 
our Monte Carlo simulations.  A small but non-negligible factor arises from 
Compton summing of $\gamma$ rays that scatter off various volumes.  This makes 
knowing the \emph{total} efficiency for $\gamma$ detection necessary, although 
the precision does not need to be very stringent because this summing with 
photopeak events is a small correction.  However, accurately quantifying the 
effects of summing is difficult due to the large volume needed for tracking 
photons that may scatter from any of the surrounding elements.  Our 
simulations where the geometry contained only detailed descriptions of the 
detectors and Mylar tape (\emph{i.e.}~neglecting the table, the floor, the 
walls, etc.) underestimates this summing effect.  Rather than attempting to 
include the many potentially important scattering surfaces, we chose to 
perform \penelope{} simulations with our geometry encased by an Aluminum 
cylinder knowing that this surely overestimates the effect.  The result of 
simulations using these two geometries is shown in 
Fig.~\ref{fig:tot-efficiency} where differences as large as a factor of two 
arise at lower energies.  We take the total efficiency to be halfway 
between these two simulated curves, with an uncertainty that spanned the 
results of both.  This results in a large uncertainty in the total efficiency 
but since the Compton summing with photopeaks is a small correction, this 
conservative estimate does not limit our determination of the $\gamma$ 
branches.

We also investigated the possible effects of $\gamma$-ray angular 
correlations, which may affect the probability of a photopeak event summing 
with the photopeak from another cascade $\gamma$.  The only cases that would 
result in a non-zero $\gamma$-$\gamma$ correlation are \mbox{$1^+\!\!\rightarrow
2^+\!\!\rightarrow0^+$} cascades; we tested both $E2$ and $M1$ transition types 
and found that independent of the transition type, $\gamma$-$\gamma$ angular 
correlations do not lead to any additional summing in our geometry.

\subsection[Q_EC and the beta efficiency]{\boldmath$Q_{EC}$ and 
  the $\beta$ Efficiency}
The mass excess of \tsups{32}Cl is obtained from our averaging 
Refs.~\cite{wrede,kankainenPRC,audiNPA} to get $\mathrm{ME}(^{32}\mathrm{Cl})=
-13334.60(57)$~keV, and $\mathrm{ME}(^{32}\mathrm{S})=-26015.535(2)$~keV is 
taken from Ref.~\cite{FSU-trap}.  Taken together, the decay energy is
$Q_{EC}=12680.9(6)$~keV.\ \ Although the fraction of events below the finite 
$E_\beta$ threshold of $40$~keV in the plastic scintillator is expected to 
have a weak dependence on the $\beta$ end-point energy, in principle the 
$\beta$ efficiency depends on $Q_\beta$.  To investigate this effect, 
\penelope{} simulations were used to determine the $\beta$ efficiency of the 
plastic scintillator.  A linear fit of the mean $\beta$ efficiency for the 
entire $\beta$ spectrum versus $\beta$ end-point energy yielded an efficiency 
that was consistent with being flat over the $4-12$-MeV range of end-points 
relevant for this decay: $\eta(Q_\beta) = [32.32(41) - 0.022(45)Q_\beta]\%$. 

\section{Data analysis}
\begin{figure}\centering
  \includegraphics[width=0.48\textwidth]{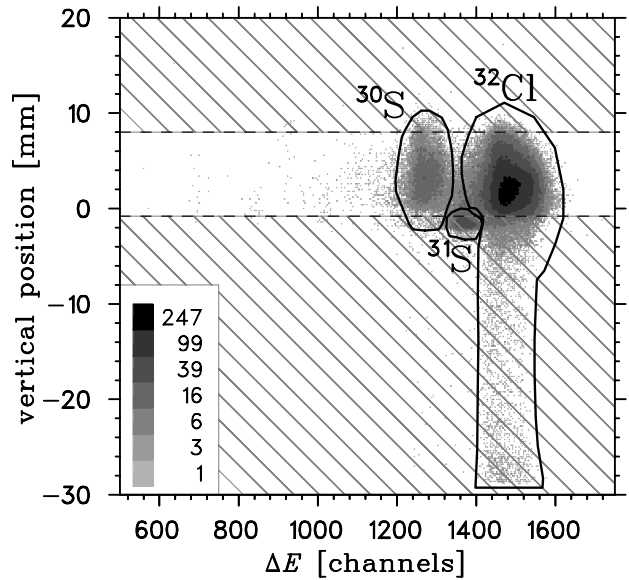}
  \caption{Identification of the ion beam at the focal plane of MARS, before 
    the purifying slits in the vertical position were moved to the closed 
    positions (indicated by the hatched areas). The region labeled 
    \tsups{32}Cl represents 86\% of all events in the spectrum, which was 
    increased to 89\% once the slits were in place.  The tails in the vertical 
    position are a result of incomplete charge collection in the 
    resistive-readout Silicon $\Delta E$ detector.\label{fig:PID}}
\end{figure}

Figure~\ref{fig:PID} shows the spectrum from the resistive-readout 
position-sensitive Silicon detector at the focal plane of MARS.\ \  
This logarithmic 2D plot shows that the most significant contaminations 
in the beam prior to our closing the purifying slits were \tsups{30}S and 
\tsups{31}S, with \tsups{32}Cl making up $\approx86\%$ of the beam at the focal 
plane of MARS.\ \  The \tsups{31}S contamination was minimized by closing the 
vertical slits at the focal plane of MARS as indicated in the figure, reducing 
it from $\approx 3\%$ to $\approx0.4\%$.  The slits had little effect at 
removing the \tsups{30}S contamination, however, as it lies in the same 
vertical band as the \tsups{32}Cl.  The purity of the beam at the focal 
plane of MARS with the vertical slits in place was $\approx89\%$.  As 
mentioned previously, the \tsups{30,31}S contaminations were further reduced 
by the degraders, which were chosen to maximize the implantation of 
\tsups{32}Cl in the centre of the Mylar tape.  The purity of the beam 
implanted in the tape was improved by another couple of percent to 
$\approx91\%$ \tsups{32}Cl with \tsups{30}S the largest contaminant at 
$\approx7.5\%$.  Note that we could have obtained a higher purity at the cost 
of reducing the rate; however, since the $\gamma$ energies of the contaminants 
do not overlap the \tsups{32}Cl lines, we chose to maximize the rate.

\begin{figure}\centering
  \includegraphics[angle=90,width=0.48\textwidth]{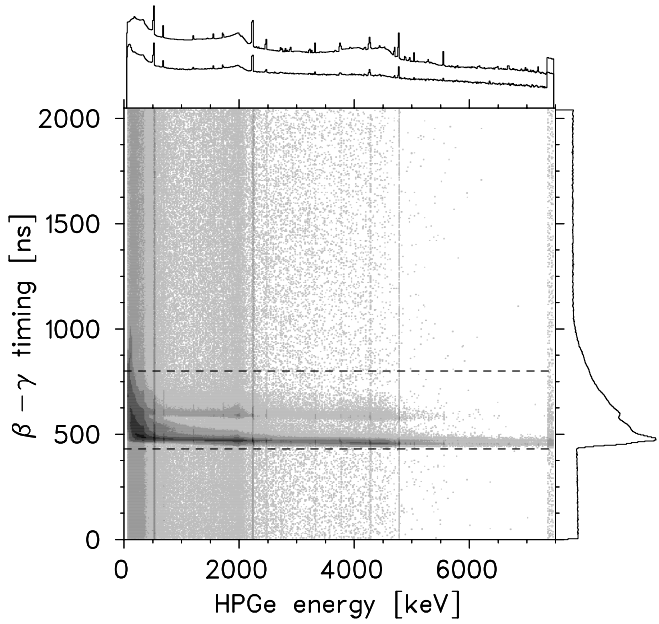}
  \caption{Logarithmic 2D plot of the $\gamma$ energy versus 
    the timing between the $\beta$ and $\gamma$.  Accidental coincident events 
    were defined to have $t_{\beta\gamma}\leq430$~ns or $t_{\beta\gamma}\geq 
    800$~ns.  The lower curve of the projection of the HPGe energy represents 
    these background events (scaled according to the cut-window size).  The 
    upper projection is the good events with $430~\mathrm{ns}<t_{\beta\gamma}<
    800$~ns.\label{fig:Egamma-vs-TOF}}
\end{figure}

The plot in Fig.~\ref{fig:Egamma-vs-TOF} shows the time difference 
between $\beta$ particles detected in the scintillator and $\gamma$-rays 
observed in the HPGe detector plotted against the $\gamma$ energy.  One clearly 
sees a strong peak at $t_\beta-t_\gamma\approx475$~ns.  In addition to the 
expected walk at low $\gamma$ energies (below 511~keV), a smaller 2\tsups{nd} 
peak around 600~ns in the timing is visible; although the 
source of this peak is not fully understood, the $\gamma$ spectrum gated on 
just it proves that these are good events.  We therefore included it in the 
analysis and defined a time gate between $430$ and $800$~ns to select 
real coincidences (the dashed horizontal lines in 
Fig.~\ref{fig:Egamma-vs-TOF}).  The rest of the timing spectrum defines another 
window which selects accidental coincidences.  The $E_\gamma$ projection 
of Fig.~\ref{fig:Egamma-vs-TOF} shows two curves:  the top one represents the 
projection of the real-coincidence window, and the lower one represents the 
accidental coincidences. The lower one, properly normalized, was then used as 
a background spectrum to be included in our fitting function as described 
below.  The real-coincidence window position and size were varied 
significantly to check for potential systematic errors.  This procedure 
yielded results with no significant sensitivity to the particular window 
used, as long as it covered the range containing all of the real coincidences. 

\begin{figure}\centering
  \includegraphics[width=0.475\textwidth]{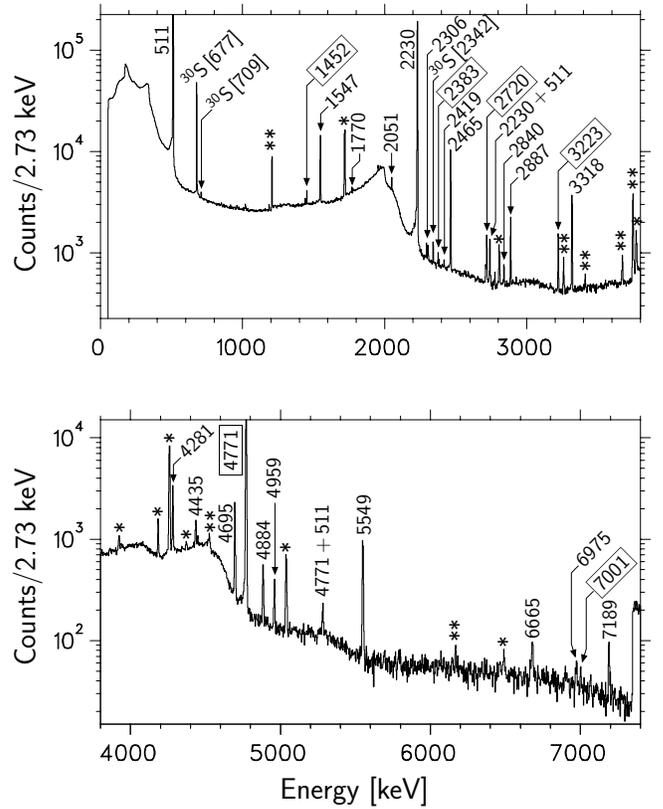}
  \caption{The $\gamma$ spectrum observed by the HPGe detector in coincidence 
    with a $\beta$.  
    Peaks are labeled with their energy, the symbol * or the symbol **, 
    referring to the 
    full energy, single-escape and double-escape peaks, respectively.  Peaks 
    not associated with \tsups{32}Cl have in addition a label indicating the 
    parent nucleus.  Decays from the 7001-keV isobaric 
    analogue state are highlighted with boxed values.  The only significant 
    background peak is at 677~keV from the \tsups{30}S contamination.
    \label{fig:spectrum}}
\end{figure}

Figure~\ref{fig:spectrum} shows the $\gamma$ spectrum observed in the HPGe 
detector in hardware coincidence with a $\beta$ signal in the scintillator.  
This is the same as the upper projection of Fig.~\ref{fig:Egamma-vs-TOF} but 
with finer binning.  Except for a strong peak at 677~keV from the decay of 
\tsups{30}S nearly every statistically significant peak is associated with 
the decay of \tsups{32}Cl. One exception is at 2776~keV where $360\pm50$
counts are seen which could not be identified based on the known levels 
in \tsups{32}S~\cite{NuclDataSheets} nor with any contaminants.  If this was, 
in fact, a product of the decay of \tsups{32}Cl, it would only represent a 
0.1\% $\gamma$-ray yield. 

\begin{figure}\centering
  \includegraphics[angle=90,width=0.48\textwidth]{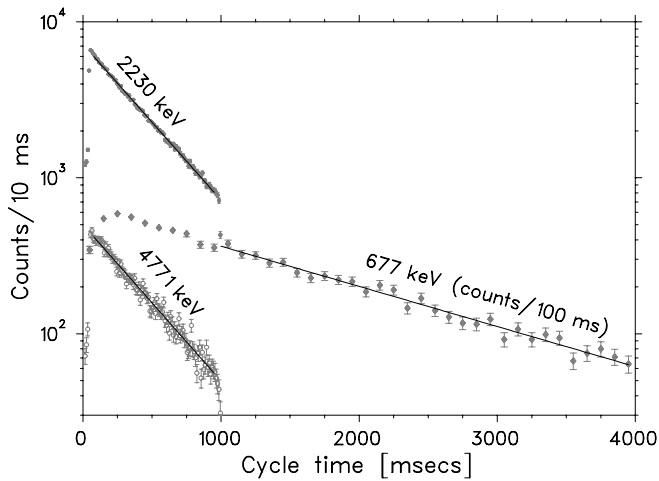}
  \caption{Number of counts as a function of time after 
    the activity was transferred by the fast tape-transport system (points) 
    and lifetime fits (solid lines).  These 
    decay spectra are gated on the two most prominent \tsups{32}Cl lines 
    (2230 and 4771~keV) as well as the dominant contaminant, 
    \tsups{30}S at 677~keV.\ \  The 
    \tsups{32}Cl peaks were fit to runs where events were counted for 1~s 
    following transfer, and the longer-lived \tsups{30}S peak used runs where 
    the count time was 4~s.\label{fig:lifetimes}}
\end{figure}

A reassuring check of the cleanliness of our data and identification of the 
\tsups{30}S contamination is shown in Fig.~\ref{fig:lifetimes}.  This shows 
a comparison of the lifetime of the two most intense peaks in the \tsups{32}Cl 
spectrum (2230 and 4771~keV) as well as the main peak (677~keV) from 
\tsups{30}S\@.  Dead time and pile-up effects were assumed to be negligible 
corrections and are not included in these half-life curves.  The fit was from 
0.060 to 0.985~s after the activity was transferred to the counting station.  
The fit lifetime of 2230-keV $\gamma$ events is $0.3012(13)$~s which is 
consistent with the $0.300(5)$~s lifetime fit from 4771-keV events.  Both 
of these results are in agreement with the accepted half-life of $0.298(1)$~s 
of Ref.~\cite{armini}.  For the 667-keV line from \tsups{30}S 
events, the Compton tail 
from higher-energy $\gamma$s from the shorter-lived \tsups{32}Cl represent 
a contamination to the \tsups{30}S curve.  In order to remove their effect, 
the minimum time included in the fit range for \tsups{30}S was 1~sec, over 
three \tsups{32}Cl half-lives.  
The fit yielded the half-life in this case to be 
$1.169(34)$~s, in good agreement with the accepted 
value of $1.1786(45)$~s from Ref.~\cite{wilsonPRC}.  Other \tsups{32}Cl 
peaks did not have enough statistics to be able to confirm the consistency 
of their $t_{1/2}$ decay curves.

\subsection[The Gamma Peak Areas]{The \boldmath$\gamma$ Peak Areas}
In order to extract peak areas we used a fitting function consisting of four 
terms:
\begin{enumerate}
\item A Gaussian, $F_\mathrm{Gauss}=\frac{1}{\sqrt{2\pi}\sigma}  
  \exp{\left(-\frac{(x-\mu)^2}{2\sigma^2}\right)}$, which corresponds to a 
  $\delta$-function at the $\gamma$ peak channel number, $\mu$, convoluted 
  with noise of width $\sigma$.  Due to the sharp rise of this peak, we 
  integrated the function over the ADC channel width, $\Delta x$, before 
  comparing to the experimentally observed histogram.  Thus the main peak 
  of our fitting function is:
  \begin{align*}
    F(x) = \frac{\frac{1}{2}\left[
        \erf\left(\frac{x-\mu+\Delta x/2}{\sqrt{2}\sigma}\right) - 
        \erf\left(\frac{x-\mu-\Delta x/2}{\sqrt{2}\sigma}\right)\right]}
    {\Delta x},
  \end{align*}
  where $\erf$ is the error function.
\item A convolution of an exponential, $F_\mathrm{exp}=\exp{\left[-\lambda(x-
      \mu)\right]}$, with a Gaussian.  Normalizing to unit area over the range 
  zero to infinity, we define a low-energy tail function corresponding to 
  incomplete charge collection as:  
  \begin{align*}
    T(x) = \frac{\lambda\ \exp\left(\frac{\lambda^2\sigma^2}{2}+\lambda(x-\mu)
      \right)\erfc\left(\frac{x-\mu+\lambda\sigma^2}{\sqrt{2}\sigma}\right)
    }{1+\erf\left(\frac{\mu}{\sqrt{2}\sigma}\right) - \exp\left(
        \frac{\lambda^2\sigma^2}{2}\!-\!\lambda\mu\right)\erfc\left(\frac{\lambda
          \sigma^2-\mu}{\sqrt{2}\sigma}\right)}.
  \end{align*}
  Here $\erfc(x)=1-\erf(x)$ is the complementary error function and $\lambda$ 
  is the decay constant of the exponential.
\item A constant background, $F_\mathrm{bkgd}=b$.
\item The background histogram from the accidental coincidences in $\beta-
  \gamma$ timing, $F_\mathrm{accid}(x)$, as described previously.
\end{enumerate}

In order that the total number of $\gamma$ signal events is a free parameter 
of the fits, we combined and normalized the terms so that our final fitting 
function was:
\begin{align}
  F_\mathrm{fit} &= a_\gamma^2 \left[(1-a_\mathrm{tail}^2)F(x) + 
    a_\mathrm{tail}^2T(x)\right] \nonumber\\ 
  &\qquad\qquad\qquad\qquad+ F_\mathrm{bkgd} + F_\mathrm{accid}(x).
  \label{eq:fit-fxn}
\end{align}
In our model, the parameters $a_\gamma$ and $a_\mathrm{tail}$ are squared so 
that neither the normalization of the photopeak area nor the relative 
amplitude of the tail due to incomplete light collection are able to converge 
to a negative value when fitting very small amplitude or non-existant 
peaks.  The total number of good $\gamma$ events in a peak is given by 
$N^\gamma=a_\gamma^2$ and the uncertainty is $\Delta N^\gamma=2a_\gamma\Delta 
a_\gamma$, where $\Delta a_\gamma$ is the statistical uncertainty 
returned from the fitting routine.  Note that by defining the fitting 
function in this way, our statistical uncertainty in the number of $\gamma$ 
events includes any correlations with other parameters of the fit.  

The $\gamma$ energies investigated included any transitions between states 
such that $E_\gamma>400$~keV.\ \ The background from the Compton edge of 
the 511~keV peak compromised the sensitivity of searching for peaks below 
this energy.  All fits were made using a \texttt{FORTRAN} code based on the 
Marquardt algorithm~\cite{marquardt} for $\chi^2$ minimization.  The program 
assumes Poisson statistics in the data (rather than Gaussian), and thus 
properly handles bins with very few counts.  The data were divided into small 
blocks with equal number of total counts and analyzed individually.  This 
procedure minimized the effects of gain variations since each data set was 
acquired over a limited range of time rather than over the whole run.  Gain 
variations determined by the change in the fit centroid value were always 
below 0.05\%; so in the end, the effects due to gain variations were 
negligible.  The number of blocks of data varied for each peak depending on 
its intensity; for example, while fitting the high-statistics 2230-keV peak 
the data were divided into 170 blocks, while the much weaker 7189-keV peak 
had all of the data summed together into one histogram before fitting.  When 
the data were divided into blocks for fitting, the areas were summed and 
uncertainties propagated to get the total number of counts.  We note that 
this number is in statistical agreement with the result obtained when fitting 
the summed histogram, although in the latter case the $\chi^2$ was worse 
due to small drifts in the gain.

\begin{figure}\centering
  \includegraphics[angle=90,width=0.485\textwidth]{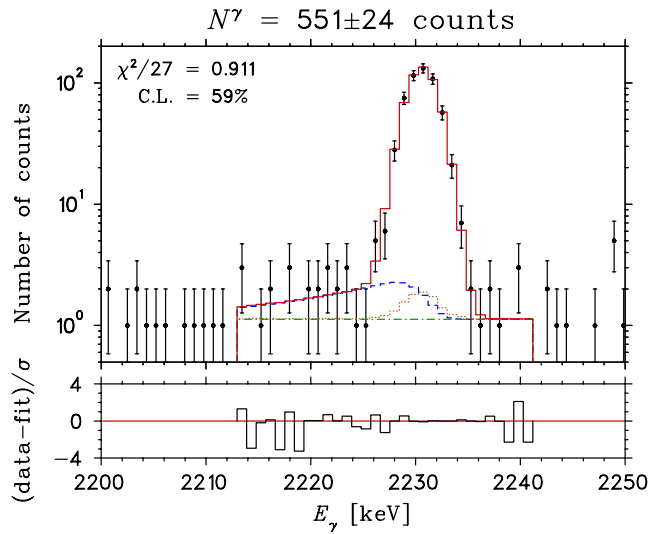}
  \caption{(Colour online) The top panel shows a typical fit, in this case to 
    the 2230-keV peak, using the fitting function of Eq.~\eqref{eq:fit-fxn}.  
    The dashed line is the tail, $T(x)$, from incomplete charge collection; 
    the dotted is the accidental coincidences, $F_\mathrm{accid}$; the dash-dot 
    line is a flat background; and the solid histogram is the total fitting 
    function.  The bottom panel shows the residual of the fit in terms of 
    $\sigma$.\label{fig:sample-fit}}
\end{figure}

Figure~\ref{fig:sample-fit} shows a typical fit to the 2230-keV peak.  In 
general the fits were excellent, with very few converging with a confidence 
level (CL) far from the ideal 50\%; the distribution of confidence levels has 
a mean of 30.8\% and a standard deviation of 24.2\%.

The energy calibration of the energy of our HPGe detector was made using: (a) 
the two most precisely known and strongest $\gamma$ de-excitations from levels 
in \tsups{32}S (2230.49(15) and 4281.51(26)~keV from Ref.~\cite{babilonPRC}); 
and (b) the three independent $\gamma$-rays from our main contaminant 
\tsups{30}S (667.01(3), 708.70(3) and 2342.2(1)~keV~\cite{NuclDataSheets}).  

\subsection[The Beta and Gamma Branches]{The \boldmath$\beta$ and $\gamma$ Branches}
Another \texttt{FORTRAN} code, also based on the Marquardt algorithm, was 
used to fit $\beta$ and $\gamma$ branches to the observed number of 
$\gamma$ events, $N^\gamma_{i,j}$, where $j$ represents the state to which 
the $i$\tsups{th} state $\gamma$ decays.  Letting $\beta_i$ represent the 
probability for the $\beta$ decay to proceed to state $i$, we normalize 
the $\beta$ branches such that
\begin{align}
  \sum\beta_i=1,
\end{align}
where $i$ is summed over all of the states considered in the analysis 
(see below).  Similarly, we normalize the $\gamma$ branch probabilities, 
$\gamma_{i,j}$, from a given \tsups{32}S excited state $i$ to lower state $j$ as
\begin{align}
  \sum_{j<i}\gamma_{i,j}=1.\label{eq:normalization-of-gamma-branch}
\end{align} 

In terms of these branches, to first order the number of observed $\gamma$ 
rays is given by:
\begin{align}
  N^\gamma_{i,j} = N_\mathrm{tot}\left[\beta_i\eta_i + 
    \sum_{k>i}\beta_k\eta_k\gamma_{k,i}\right]\gamma_{i,j}\epsilon_{i,j},
  \label{eq:N_gamma}
\end{align}
where the first term in the brackets arises from the $\beta$ transition 
directly to state $i$, and the second term represents the feeding from $\beta$ 
transitions to higher levels that $\gamma$ decay to state $i$.  Here 
$N_\mathrm{tot}$ is the total number of decays (a free parameter in the fits 
since the total number of $\beta$'s was not precisely measured), 
$\epsilon_{i,j}$ is the photopeak efficiency of the HPGe detector at an energy 
$E_{i,j}$, and $\eta_i$ is the efficiency of the plastic scintillator to 
observe a $\beta$ with an end-point energy $Q_i$.  Small corrections to 
Eq.~\eqref{eq:N_gamma} (included in the analysis but omitted here for 
clarity) are required to account for (a) summing with cascade $\gamma$s from 
above and below (which requires the \emph{total} efficiency of the HPGe), and 
(b) summing with the $511$ annihilation radiation since this is a $\beta^+$ 
decay.

The nine lowest $(0,1,2)^+$ states observed by D\'etraz \emph{et 
al.}~\cite{detraz} and Anderson \emph{et al.}~\cite{anderson} along with the 
$(1.0^{+0.2}_{-0.5})\%$ ground state branch from Armini 
\emph{et al.}~\cite{armini} represent most of the $\beta$ yield; however 
little is known above $7.2$~MeV of excitation energy.  Some $\beta$-delayed 
proton and alpha decays have been observed~\cite{honkanenNPA} to states above 
8.7~MeV with particle energies as low as 762(5)~keV, but this still leaves a 
$1.5$~MeV window of $Q$-value in which no $\beta$ transitions have been 
identified.  This suggests that there is no appreciable $\beta$ feeding in this 
energy region; however, it does not rule out the possibility 
of a large number of weak $\beta$ transitions.  Each of these transitions may 
be too small to be detected individually, but could cumulatively contribute 
a total $\beta$ strength of up to a few per cent.   This ``Pandemonium 
effect,'' originally pointed out in Ref.~\cite{hardy:pandemonium}, was raised 
again recently~\cite{HT:02} with regard to superallowed $\beta$ decay in 
$p,f$-shell nuclei.  Following the approach advocated in these references, we 
use a shell-model calculation to compute the weak $\beta$ branches and include 
its predicted strength in the analysis.  The model space used is the full 
$s,d$ shell with the effective interactions \usd{} of Wildenthal~\cite{USD} 
and the two more recent updates \usda{} and \usdb{} of Brown and 
Richter~\cite{USDAB}. 

We include in our analysis of the branches and yields a total of $51$ excited 
states in \tsups{32}S.\ \ Our shell-model calculation correctly predicts all 
of the $9$ lowest $(0,1,2)^+$ states with $E_x<7.2$~MeV reported in D\'etraz 
\emph{et al.}~\cite{detraz}.   We find that the RMS deviations of the 
shell-model calculation from the known excitation 
energies~\cite{NuclDataSheets} are quite 
good: 120~keV (\usd), 209~keV (\usda) and 172~keV (\usdb).  This is a 
gratifying indication that the shell model is performing well in this 
$s,d$-shell nucleus.  Even though selection rules prohibit $\beta$ decays to 
the six lowest $(3,4)^\pm$ states, they are included in the analysis when 
accounting for $\gamma$-ray de-excitations. The shell-model calculations 
identify approximately $40$ $\beta$ transitions to states whose excitation 
energies in \tsups{32}S lies between $7.485$ and $\approx11.8$~MeV.\ \ 
Unfortunately, the high density of states in this energy range makes a 
state-by-state comparison difficult, especially for the $2^+$ states.  Based 
on the good correspondence of excitation energies and $\gamma$ de-excitation 
branches, we are able to identify $6$ of the shell-model states in this region 
with ones in the ENSDF Data Tables.  None of the other $30$ shell-model states 
individually has a $\beta$-transition strength greater than $0.3\%$, but 
cumulatively they sum to $0.50\%$ in the \usd{}, $0.69\%$ in the \usda{}, and 
$0.55\%$ in the \usdb{} calculations.  We include these weak $\beta$ strengths 
and de-excitation $\gamma$ rays predicted by the shell model in our overall 
analysis.  

In the analysis, a $\beta$ branch could be deduced as long as there is at least 
one decay $\gamma$ ray lying within the 7.35~MeV energy range of our HPGe.
The ground-state branch and higher excitation-energy shell-model-state 
branches that were not observed in this experiment were included in the 
analysis as missing strength.  For the ground state, we take the branch to 
be $(1.0^{+0.2}_{-0.5})\%$ as determined by Armini \emph{et al.}~\cite{armini}, 
and the combination of all the unseen shell-model states at energies above 
7.2~MeV is taken to be the average of the \usd, \usda{} and \usdb{} 
calculations with an uncertainty that spans the variation: $(0.60\pm0.10)\%$.  
The $\alpha$-particle and proton-emitting states in the $8.7$ to $11.1$~MeV 
excitation energy range reported by Honkanen \emph{et al.}~\cite{honkanenNPA} 
are not separately included because their summed $\beta$ strength of 
$(0.080\pm0.005)\%$ is significantly less than and no doubt already included 
in the missing strength predicted by the shell model.

\section{Results\label{section:Results}}
\subsection{Experimental Results}
The excitation energies in \tsups{32}S, the $\beta$-decay branches and the 
$\log{ft}$ values determined in this work are shown in 
Table~\ref{table:beta-branches}. The states up to 7.2~MeV each have multiple 
$\gamma$ rays which were observed in this work.  Thus the $\gamma$ branches 
in these cases could be fit using the procedure described earlier.  
Furthermore, the fitting routine allowed us to treat the 
excitation energies of these states as free parameters, and fit the observed 
lines such that the $E_x$ values minimized the $\chi^2$ of the calculated  
$\gamma$ energies.  For the highest three energy levels listed in 
Table~\ref{table:beta-branches}, only one de-excitation $\gamma$ ray was 
observed; therefore, in these cases the $\gamma$ branches had to be taken 
from previous work~\cite{NuclDataSheets}, and the excitation energy had to be 
calculated solely on the one calibrated $\gamma$ ray that was observed.

In Table~\ref{table:gamma-branches} we list the $\gamma$ branches for the 
states which have multiple $\gamma$ rays within our $7.35$~MeV energy range.  
The $\gamma$ energy in this table is calculated by first averaging the 
ENSDF~\cite{NuclDataSheets} $E_x$'s with our own (the first two columns of 
Table~\ref{table:beta-branches}) and then calculating the Doppler-corrected 
$E_\gamma$ based on the new excitation energies.

\begin{table*}
  \caption{Excitation energies in \tsups{32}S, deduced $\beta$ branches and the 
    comparative half-lives of the decay of \tsups{32}Cl. The $0^+$ spin assignment 
    of the $7637$- and $7921$-keV levels are determined from a comparison with our 
    shell-model calculation. Not listed are the weakly-fed states between 
    $7.4-11.8$~MeV predicted by the shell-model to have a total branch of 
    $(0.60\pm0.10)\%$. \label{table:beta-branches}}
  \begin{tabular}{D{.}{.}{2}@{$\pm$}D{.}{.}{3}
      D{.}{.}{1}@{$\pm$}D{.}{.}{2}
      D{.}{.}{2}@{$\pm$}D{.}{.}{3}
      ccc
      D{.}{.}{4.1}@{$\pm$}D{.}{.}{3}
      D{.}{.}{3}@{$\pm$}l
      c
      D{.}{.}{2.2}@{$\pm$}D{.}{.}{4}
      r@{.}l}
    \hline
    \hline
    \multicolumn{18}{c}{}\\[-0.85em]
    \multicolumn{6}{c}{$E_x$ in \tsups{32}S [keV]} && && \multicolumn{4}{c}{$\beta$ branch [\%]} && \multicolumn{4}{c}{$\log ft$} \\[-0.5em]
    \multicolumn{6}{c}{} & \hspace*{0.5em} & $J^\pi_n, T$ &\hspace*{0.5em}& \multicolumn{4}{c}{} &\hspace*{1em}& \multicolumn{4}{c}{} \\[-0.5em]
    \multicolumn{2}{c}{ENSDF$^a$} & \multicolumn{2}{c}{This work} & \multicolumn{2}{c}{Average\ \ \ } && && \multicolumn{2}{c}{D\'etraz \emph{et al.}$^b$} & \multicolumn{2}{c}{This work} 
    && \multicolumn{2}{c}{D\'etraz \emph{et al.}$^b$} & \multicolumn{2}{c}{\ \ This work\ \ \ \ } \\
    \hline
    \multicolumn{18}{c}{\ }\\[-0.8em]
    9207.55 & 0.71   & 9204.7&1.7             & 9207.1 &0.7   &&$1^+_5,1$&&\multicolumn{2}{c}{--}&  0.22&$0.04^{+0.02}_{-0.09}$       && \multicolumn{2}{c}{--}& 4 &3$_{-0.1}^{+0.4}$         \\[0.25em]
    8125.40 & 0.20   & \multicolumn{2}{c}{--} & 8125.40&0.20  &&$1^+_4,1$&&\multicolumn{2}{c}{--}&  \multicolumn{2}{c}{\ $<0.19$}  && \multicolumn{2}{c}{--}&$>5$&$0$                     \\[0.25em]
    7921.0  & 1.0    & 7924.3&1.8             & 7921.8 &0.9   &&$0^+_5,0$&&\multicolumn{2}{c}{--}&  0.033&$0.012^{+0.010}_{-0.004}$    && \multicolumn{2}{c}{--}& 5 &9$_{-0.1}^{+0.3}$        \\[0.25em]
    7637.0  & 1.0    & \multicolumn{2}{c}{--} & 7637.0 &1.0   &&$0^+_4,0$&&\multicolumn{2}{c}{--}&  \multicolumn{2}{c}{\ $<0.026$} && \multicolumn{2}{c}{--}&$>6$&$2$                     \\[0.25em]
    7535.7  & 1.0    & 7535.3&1.8             & 7535.6 &0.9   &&$0^+_3,1$&&\multicolumn{2}{c}{--}&  0.185&$0.017^{+0.019}_{-0.003}$    && \multicolumn{2}{c}{--}& 5 &34$_{-0.05}^{+0.04}$     \\[0.25em]
    7190.1  & 1.5    & 7190.5&1.6             & 7190.3 &1.1   && $1^+_3,0$ &&   0.9&0.1  &  0.62&$0.05\!\pm\!0.01$                 && 4.90&0.10             & 4 &98$_{-0.03}^{+0.04}$    \\[0.25em]
    7115.3  & 1.0    & 7114.7&1.5             & 7115.1 &0.8   && $2^+_5,1$ &&   0.5&0.2  &  0.62&$0.04^{+0.00}_{-0.02}$                && 5.1&0.2               & 5 &01$\pm$0.03             \\[0.25em]
    7001.4  & 0.4    & 7001.0&1.3             & 7001.4 &0.4   && $1^+_2,1$ &&  20.5&2.0  & 22.47&$0.13^{+0.16}_{-0.12}$                && 3.52&0.04             & 3 &500$\pm$0.004           \\[0.25em]
    6666.1  & 1.0    & 6665.4&1.4             & 6665.9 &0.8   && $2^+_4,0$ &&   1.8&0.5  &  2.09&$0.07^{+0.01}_{-0.03}$                && 4.72&0.12             & 4 &671$^{+0.016}_{-0.014}$ \\[0.25em]
    5548.5  & 1.0    & 5548.3&1.1             & 5548.4 &0.8   && $2^+_3,0$ &&   4.1&0.5  &  3.83&$0.07\!\pm\!0.09$                  && 4.77&0.06            &  4 &816$^{+0.014}_{-0.012}$ \\[0.25em]
    4695.3 & 0.4     & 4695.5&0.9             & 4695.3 &0.4   && $1^+_1,0$ &&   6.8&0.8  &  6.10&$0.08^{+0.03}_{-0.04}$                && 4.81&0.05             & 4 &880$^{+0.007}_{-0.006}$ \\[0.25em]
    4281.8 & 0.3     & 4281.9&0.8             & 4281.81&0.28  && $2^+_2,0$ &&   3.1&0.4  &  2.18&$0.08\!\pm\!0.02$                  && 5.45&0.07            &  5 &444$^{+0.016}_{-0.015}$ \\[0.25em]
    3778.4  & 1.0    & 3778.1&0.9             & 3778.3 &0.7   && $0^+_2,0$ &&   2.6&0.8  &  0.95&$0.07^{+0.01}_{-0.05}$                && 5.48&0.13             & 5 &94$_{-0.03}^{+0.04}$    \\[0.25em]
    2230.57 & 0.15   & 2230.4&0.5             & 2230.56&0.14  && $2^+_1,0$ &&    60&4    & 59.08&$0.18^{+0.39}_{-0.26}$                && 4.49&0.04             & 4 &516$^{+0.002}_{-0.003}$ \\[0.25em]
    \multicolumn{6}{c}{ground state\ \ }                      && $0^+_1,0$ && \multicolumn{4}{c}{not measured$^c$}                  && \multicolumn{4}{c}{not measured$^d$}       \\[0.25em]
    \hline
    \hline
    \multicolumn{18}{l}{}\\[-0.75em]
    \multicolumn{18}{l}{$^a$Ref.~\cite{NuclDataSheets}.}\\
    \multicolumn{18}{l}{$^b$Ref.~\cite{detraz}.}\\
    \multicolumn{18}{l}{$^c$Fixed to $1.0^{+0.2}_{-0.5}\%$ from Armini, \emph{et al.}~\cite{armini}.}\\
    \multicolumn{18}{l}{$^d$Calculated to be $6.7^{+0.2}_{-0.1}$.}\\
  \end{tabular}\\\raggedright
\end{table*}

\begin{table*}
  \caption{$\gamma$-decay branches, in percent, for excited states in 
    \tsups{32}S and comparison to the currently accepted values in 
    ENSDF~\cite{NuclDataSheets}.  The 2230.48(14)~keV $2^+_1$ transition 
    to the $0^+_1$ ground state is assumed to be 100\% since there are no 
    other known levels to which the $2^+_1$ state may decay.  The energies 
    of the $\gamma$ rays are calculated based on the adopted energy levels, 
    \emph{i.e.}~the average of our work and ENSDF, the third column in 
    Table~\ref{table:beta-branches} and shown in 
    Fig.~\ref{fig:decay_scheme}. \label{table:gamma-branches}}
  \begin{tabular}{rlcr@{}lcr@{.}lcrlcr@{}lcr@{.}l} 
    \hline
    \hline
    \multicolumn{14}{c}{\ } \\[-0.8em]
    \multicolumn{1}{c}{Transition} &
    \multicolumn{1}{c}{$E_\gamma$\,[keV]} &
    \multicolumn{4}{c}{ENSDF} & 
    \multicolumn{2}{c}{Present work}                                 &\hspace*{1em}&
    \multicolumn{1}{c}{Transition} &
    \multicolumn{1}{c}{$E_\gamma$\,[keV]} & 
    \multicolumn{4}{c}{ENSDF} & 
    \multicolumn{2}{c}{Present work}\\
    \hline
    \multicolumn{14}{c}{\ } \\[-0.8em]
 $1^+_3\rightarrow2^+_3$ & $1641.8\pm1.3$ &\ &   &      &\ &  $<6$&$7$                      && $2^+_4\rightarrow2^+_3$ & $1117.4\pm1.1$ &\ &    &         &\ &$<1$&$1$                         \\
                 $1^+_1$ & $2494.9\pm1.1$ &&$<42$&      &&  $2$&$6\!\pm\!2.1^{+0.1}_{-0.7}$    &&                $1^+_1$ & $1970.5\pm0.9$ && $14$&$\pm2$     &&  $7$&$3\!\pm\!1.8^{+0.0}_{-0.2}$    \\
                 $2^+_2$ & $2908.4\pm1.1$ &&$<35$&      &&  $<3$&$4$                        &&                $2^+_2$ & $2384.0\pm0.9$ && $<7$&           &&  $3$&$7\!\pm\!0.9^{+0.0}_{-0.3}$    \\
                 $0^+_2$ & $3411.9\pm1.3$ &&$<55$&        &&  $19$&$5\!\pm\!2.7^{+0.2}_{-1.1}$ &&                $0^+_2$ & $2887.5\pm1.0$ && $49$&$\pm5$     && $46$&$7\!\pm\!1.6^{+0.7}_{-0.2}$    \\
                 $2^+_1$ & $4959.3\pm1.1$ && $59$&$\pm12$ &&  $50$&$8\!\pm\!3.8^{+0.8}_{-1.4}$ &&                $2^+_1$ & $4435.0\pm0.8$ && $37$&$\pm4$     && $39$&$6\!\pm\!1.6^{+0.4}_{-0.7}$    \\
                 $0^+_1$ & $7189.4\pm1.1$ && $41$&$\pm12$ &&  $27$&$1\!\pm\!2.9^{+2.0}_{-0.8}$ &&                $0^+_1$ & $6665.1\pm0.8$ && $<3$&           &&  $2$&$3\!\pm\!0.8^{+0.3}_{-0.4}$    \\
 \multicolumn{14}{c}{\ }\\[-0.8em]
 \multicolumn{8}{c}{\ }                                                                    && $2^+_3\rightarrow1^+_1$ &\ \ $853.1\pm0.8$&&     &          &&  $0$&$65\!\pm\!0.20\!\pm\!0.03$  \\
 \multicolumn{8}{c}{\ }                                                                    &&                 $2^+_2$ & $1266.6\pm0.8$ && $<1$&           &&  $0$&$86\!\pm\!0.30^{+0.12}_{-0.02}$\\
 $2^+_5\rightarrow2^+_3$ & $1566.6\pm1.1$ &&     &         && $<4$&$6$                      &&                 $0^+_2$ & $1770.1\pm1.0$ && $<1$&           &&  $3$&$3\!\pm\!0.6\!\pm\!0.1$    \\
                $1^+_1$ & $2419.7\pm0.9$ &&\ \ $9$&$\pm1$ &&  $9$&$0\!\pm\!2.0^{+0.1}_{-1.0}$  &&                 $2^+_1$ & $3317.7\pm0.8$ && $60$&.0$\pm$1.5 && $59$&$2\!\pm\!0.8^{+1.1}_{-0.9}$   \\   
                $2^+_2$ & $2833.2\pm0.9$ &&\ \ $3$&$\pm1$ &&  $3$&$0\!\pm\!2.0^{+0.6}_{-0.4}$  &&                 $0^+_1$ & $5547.9\pm0.7$ && $40$&.0$\pm$1.5 && $36$&$1\!\pm\!0.7^{+1.0}_{-1.1}$   \\    
                $0^+_2$ & $3336.7\pm1.1$ &&\ \ $3$&$\pm2$ &&  $5$&$8\!\pm\!2.2^{+0.0}_{-0.3}$  && \multicolumn{8}{c}{\ }                                                                        \\
                $2^+_1$ & $4884.1\pm0.8$ && $83$&$\pm2$   && $79$&$3\!\pm\!3.9^{+1.2}_{-0.5}$  && $1^+_1\rightarrow2^+_2$ &\ \ $413.5\pm0.5$ && $<0$&$.6$        && $<0$&$28$                   \\
                $0^+_1$ & $7114.3\pm0.8$ &&\ \ $2$&.9$\pm$0.5 && $<4$&$6$                   &&                $0^+_2$ &\ \ $917.1\pm0.8$ && $<0$&$.4$        &&  $0$&$50\!\pm\!0.14^{+0.01}_{-0.04}$\\   
 \multicolumn{8}{c}{\ }                                                                    &&                 $2^+_1$ & $2464.7\pm0.4$ && $61$&.0$\pm$1.0    &&  $63$&$3\!\pm\!0.5^{+0.2}_{-0.1}$\\
 \multicolumn{8}{c}{\ }                                                                    &&                 $0^+_1$ & $4695.0\pm0.4$ && $39$&.0$\pm$1.0    &&  $36$&$2\!\pm\!0.5^{+0.1}_{-0.2}$\\
 \multicolumn{8}{c}{\ }                                                                    && \multicolumn{8}{c}{\ }                                                                         \\
 $1^+_2\rightarrow2^+_3$ & $1452.9\pm0.8$ &&       &    &&  $1$&$23\!\pm\!0.08\!\pm\!0.01$  && $2^+_2\rightarrow0^+_2$ &\ \ $503.5\pm0.7$ &&$<0$&$.3$         && $<1$&$5$                      \\
                 $1^+_1$ & $2305.9\pm0.5$ && $<1$ &     &&  $0$&$61\!\pm\!0.10\!\pm\!0.03$  &&                $2^+_1$ & $2051.2\pm0.3$   && $13$&.0$\pm$0.5  && $16$&$1\!\pm\!0.9^{+0.1}_{-0.4}$ \\
                 $2^+_2$ & $2719.4\pm0.5$ && $<2$ &     &&  $2$&$37\!\pm\!0.08^{+0.01}_{-0.07}$ &&                 $0^+_1$ & $4281.5\pm0.3$   && $87$&.0$\pm$0.5 && $83$&$5\!\pm\!1.1^{+0.3}_{-0.1}$ \\
                 $0^+_2$ & $3222.9\pm0.8$ && $9$&$\pm5$ &&  $3$&$92\!\pm\!0.12^{+0.05}_{-0.02}$ && \multicolumn{8}{c}{\ }                                                                        \\
                 $2^+_1$ & $4770.4\pm0.4$ && $91$&      && $91$&$63\!\pm\!0.19^{+0.06}_{-0.04}$ && $0^+_2\rightarrow2^+_1$ & $1547.7\pm0.7$ && $99$&$.965$       && $98$&$6\!\pm\!0.8^{+0.2}_{-0.0}$ \\
                 $0^+_1$ & $7000.5\pm0.4$ && $<2$ &     &&  $0$&$25\!\pm\!0.07\!\pm\!0.01$  &&                $0^+_1$ & $3778.0\pm0.7$ && $0$&.035$\pm$0.006 && $1$&$4\!\pm\!0.8^{+0.0}_{-0.2}$\\[0.25em]
    \hline
    \hline
  \end{tabular}
\end{table*}

\begin{figure*}
  \includegraphics[angle=0,width=0.975\textwidth]{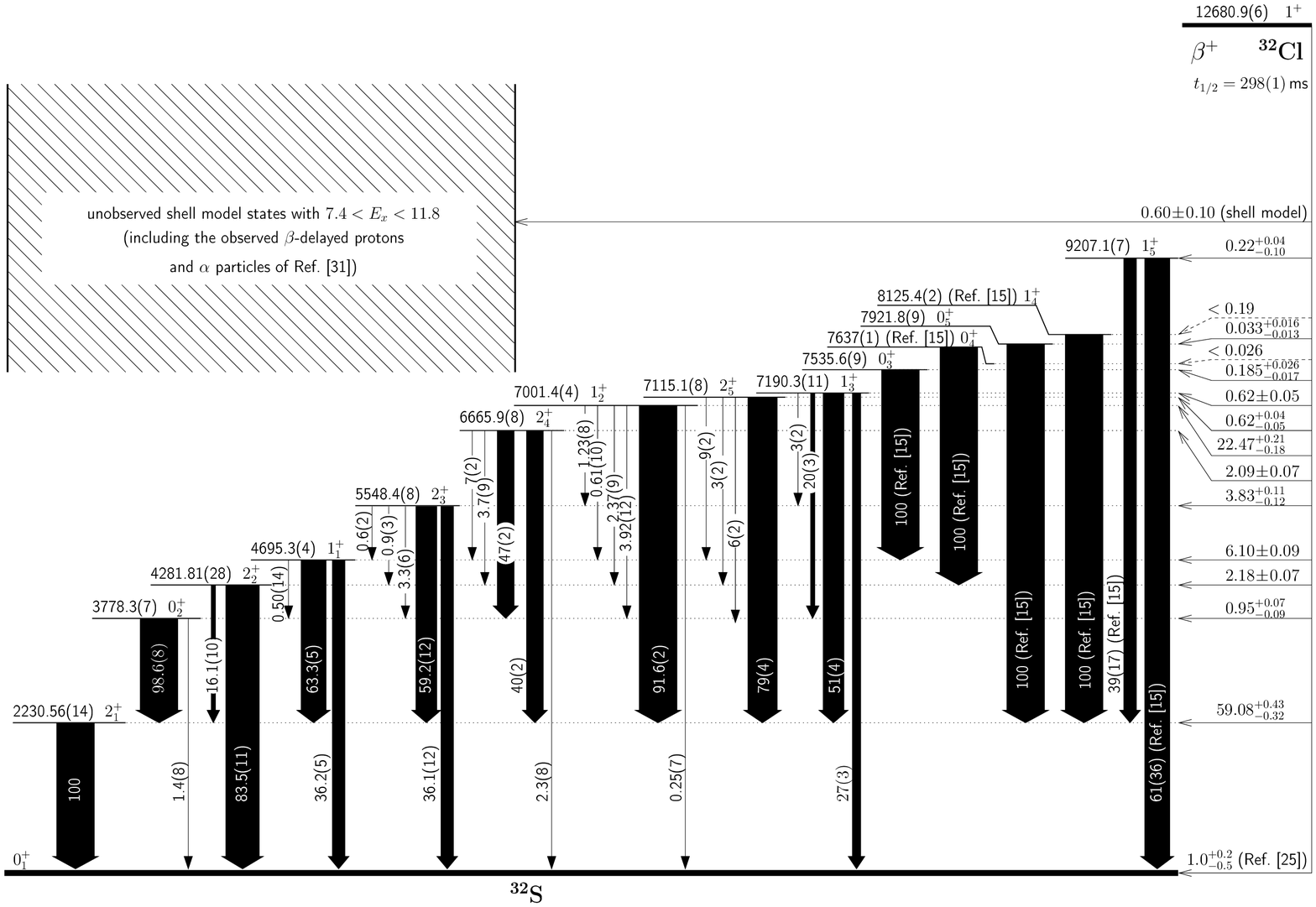}
  \caption{$\beta^+$ decay scheme for $^{32}$Cl, summarizing the 
      $\gamma$ and $\beta$ branches in Tables~\ref{table:beta-branches} 
      and~\ref{table:gamma-branches} determined from this work.  All branches 
      are expressed in percent.  The excitation energies are the weighted 
      average of the present work and the accepted values from 
      ENSDF~\cite{NuclDataSheets} (third column of 
      Table~\ref{table:beta-branches}).\label{fig:decay_scheme}}
\end{figure*}

Figure~\ref{fig:decay_scheme} graphically depicts the results listed in 
Tables~\ref{table:beta-branches} and~\ref{table:gamma-branches}.  The figure 
also includes within it the references to previous results which were used 
in the analysis.  For example, the $\gamma$ ray from the $1_5^+$ state at 
9206~keV was outside the energy range of our HPGe, so we took its $\gamma$ 
de-excitation branching ratio from the ENSDF Data Tables in order to fit the 
$\beta$ branch based on the observed $\gamma$ ray to the $2^+_1$ first excited 
state.

In both Tables~\ref{table:beta-branches} and~\ref{table:gamma-branches}, the 
first uncertainty in the branches is statistical and the second is systematic.  
The sources of systematic error considered include: the cuts made on the 
data (the $\beta$/HI ratio, the $\beta$--$\gamma$ timing windows for real 
versus accidental coincidences, and the start/stop of the counting time 
following transfer of the activity); the uncertainty in the $\gamma$ photopeak 
areas and total efficiencies; the $\beta$ efficiency; the effective interaction 
of the shell model used for weakly-fed states above 7.2~MeV; 
including/excluding the ENSDF $\gamma$ branches for higher levels (and varying 
them by their uncertainties when included); and the $^{+0.2}_{-0.5}\%$ 
uncertainty in the ground state branch.  For the $\gamma$ branches in 
Table~\ref{table:gamma-branches}, the sum of the quoted probabilities for 
decay from a given excited state in \tsups{32}S is not necessarily 100\%, 
\emph{i.e.}\ contrary to Eq.~\eqref{eq:normalization-of-gamma-branch}, it 
may seem that $\sum_{j<i}\gamma_{i,j}\leq1$; this is because of possible--but 
not statistically significant--peaks where the data only allows us to place 
limits on the branch. As with the $\gamma$ yields discussed below, a branch 
is quoted only if the area of a $\gamma$-ray peak in the HPGe was larger than 
its total uncertainty. If this condition is not satisfied, we instead quote 
an upper limit on the branch at the 90\% C.L.\ \ For example, the $2^+_5$ 
state at 7115~keV has four statistically significant branches which have a 
combined probability of only $97.1\pm5.3^{+1.3}_{-1.2}$\%.  For the other two 
branches, the fit converged to results that were consistent with zero and for 
which only upper limits may be quoted:  $1.2\!\pm\!2.3^{+0}_{-0.5}\%$ ($2^+_3$) 
and $1.7\!\pm\!2.0^{+0.6}_{-0.1}\%$ ($0^+_1$).  All together, including these 
statistically insignificant transition strengths, the total probability 
\emph{is} 100\%; the ``missing'' 2.9\% in Table~\ref{table:gamma-branches} 
for the decay from the $2^+_5$ state is potentially within transitions which 
are below our detection sensitivity.  Note, however, that the $1$-$\sigma$ 
upper limit on the sum of the statistically significant branches includes 
100\%.

With the $\beta$ and $\gamma$ branches established, we are able to calculate 
the $\gamma$ yields.  The reason we present the yields after the discussion of 
the branches is because of the small summing corrections which depend on the 
branching ratios.  The results are listed in Table~\ref{table:gamma-yields} 
where we again compare the excitation energies to ENSDF~\cite{NuclDataSheets} and the 
yields to the work of D\'etraz \emph{et al.}~\cite{detraz}.

\begin{table*}
  %
  %
  %
  \caption{Yields of $\gamma$-rays following the $\beta$ decay of 
    \tsups{32}Cl.  All values are in percent, and limits correspond to the 
    90\% confidence level. The overall normalization is fixed such that the 
    transition to the ground state is consistent with $1.0^{+0.2}_{-0.5}\%$ as 
    reported by Armini, \emph{et al.}~\cite{armini} and includes the 
    $(0.60\pm0.10)\%$ unobserved, weakly-fed shell-model states in the 
    energy range of $7.4-11.8$~MeV.\label{table:gamma-yields}}
  \begin{tabular}{D{.}{.}{2}@{$\pm$}D{.}{.}{2}D{.}{.}{1}@{$\pm$}D{.}{.}{1}cllclrcD{.}{.}{2}@{$\pm$}D{.}{.}{4}D{.}{.}{9}}
    \hline
    \hline
    \multicolumn{14}{c}{\ }\\[-0.75em]
    \multicolumn{4}{c}{$E_\gamma$ [keV]} &\hspace*{2em}& &&&& &\hspace*{2em}& \multicolumn{3}{c}{$\gamma$ yield [\%]}\\[-0.4em]
    \multicolumn{4}{c}{} &&  \multicolumn{5}{c}{Transition}  && \multicolumn{3}{c}{} \\[-0.4em]
    \multicolumn{2}{c}{ENSDF$^a$} & \multicolumn{2}{c}{This work} && &&&& && \multicolumn{2}{c}{D\'etraz \emph{et al.}$^b$} & 
    \multicolumn{1}{c}{This work}\\
    \hline
    \multicolumn{14}{c}{} \\[-0.75em]
    7189.2&1.5  & 7189.8&1.2           && $1^+_3$&$(7190)$&$\rightarrow$&$0^+_1$&$(\mathrm{g.s.})$ && 0.41&0.10           &  0.169^{+0.024}_{-0.020}  \\
    7114.5&1.0  & \multicolumn{2}{c}{} && $2^+_5$&$(7115)$&$\rightarrow$&$0^+_1$&$(\mathrm{g.s.})$ &&\multicolumn{2}{c}{} & <0.029                    \\ 
    7000.6&0.4  & 7001.4&1.6           && $1^+_2$&$(7001)$&$\rightarrow$&$0^+_1$&$(\mathrm{g.s.})$ &&\multicolumn{2}{c}{} &  0.057\!\pm\!0.016        \\
    6976.2&0.7  & 6973.5&1.3           && $1^+_5$&$(9208)$&$\rightarrow$&$2^+_1$&$(2230)$          &&\multicolumn{2}{c}{} &  0.098\pm0.018            \\ 
    6665.4&1.0  & 6665.8&2.1           && $2^+_4$&$(6666)$&$\rightarrow$&$0^+_1$&$(\mathrm{g.s.})$ &&\multicolumn{2}{c}{} &  0.048^{+0.018}_{-0.019}  \\ 
    5894.2&0.2  & \multicolumn{2}{c}{} && $1^+_4$&$(8125)$&$\rightarrow$&$2^+_1$&$(2230)$          &&\multicolumn{2}{c}{} &  <0.027                   \\ 
    5689.9&1.0  & 5693.3&1.3           && $0^+_5$&$(7921)$&$\rightarrow$&$2^+_1$&$(2230)$          &&\multicolumn{2}{c}{} &  0.033^{+0.014}_{-0.013}  \\
    5548.0&1.0  & 5548.9&0.9           && $2^+_3$&$(5549)$&$\rightarrow$&$0^+_1$&$(\mathrm{g.s.})$ && 1.6&0.3             &  1.50^{+0.08}_{-0.09}     \\
    4959.1&1.5  & 4959.6&0.8           && $1^+_3$&$(7190)$&$\rightarrow$&$2^+_1$&$(2230)$          &&\multicolumn{2}{c}{} &  0.32\!\pm\!0.04          \\
    4884.3&1.0  & 4883.7&0.8           && $2^+_5$&$(7115)$&$\rightarrow$&$2^+_1$&$(2230)$          && 0.45&0.20           &  0.504^{+0.031}_{-0.032}  \\
    4770.4&0.4  & 4770.8&0.8           && $1^+_2$&$(7001)$&$\rightarrow$&$2^+_1$&$(2230)$          && 20.5&2.0            & 20.62^{+0.20}_{-0.17}     \\
    4694.9&0.4  & 4695.6&0.8           && $1^+_1$&$(4695)$&$\rightarrow$&$0^+_1$&$(\mathrm{g.s.})$ && 2.8&0.6             &  2.42\!\pm\!0.05          \\
    4435.2&1.0  & 4435.5&0.8           && $2^+_4$&$(6666)$&$\rightarrow$&$2^+_1$&$(2230)$          && 0.8&0.2             &  0.83\!\pm\!0.06          \\
    4281.5&0.3  & 4282.0&0.7           && $2^+_2$&$(4282)$&$\rightarrow$&$0^+_1$&$(\mathrm{g.s.})$ && 2.6&0.1             &  2.42\!\pm\!0.06          \\
    3778.2&1.0  & 3777  &4             && $0^+_2$&$(3778)$&$\rightarrow$&$0^+_1$&$(\mathrm{g.s.})$ &&\multicolumn{2}{c}{} &  0.044\!\pm\!0.025        \\ 
    3411.5&1.8  & 3412.2&0.7           && $1^+_3$&$(7190)$&$\rightarrow$&$0^+_2$&$(3778)$          &&\multicolumn{2}{c}{} &  0.122\!\pm\!0.019        \\
    3355.0&1.0  & \multicolumn{2}{c}{} && $0^+_4$&$(7637)$&$\rightarrow$&$2^+_2$&$(4282)$          &&\multicolumn{2}{c}{} &  <0.026                   \\ 
    3336.7&1.4  & 3339.7&1.2           && $2^+_5$&$(7115)$&$\rightarrow$&$0^+_2$&$(3778)$          &&\multicolumn{2}{c}{} &  0.037\!\pm\!0.015        \\ 
    3317.7&1.0  & 3317.9&0.6           && $2^+_3$&$(5549)$&$\rightarrow$&$2^+_1$&$(2230)$          && 2.5&0.4             &  2.46\!\pm\!0.05          \\
    3222.8&1.1  & 3222.4&0.6           && $1^+_2$&$(7001)$&$\rightarrow$&$0^+_2$&$(3778)$          &&\multicolumn{2}{c}{} &  0.881^{+0.029}_{-0.027}  \\
    2908.2&1.5  & \multicolumn{2}{c}{} && $1^+_3$&$(7190)$&$\rightarrow$&$2^+_2$&$(4282)$          &&\multicolumn{2}{c}{} &  <0.022                   \\ 
    2887.6&1.4  & 2887.0&0.5           && $2^+_4$&$(6666)$&$\rightarrow$&$0^+_2$&$(3778)$          && 1.0&0.4             &  0.976^{+0.028}_{-0.025}  \\
    2840.3&1.1  & 2839.7&0.5           && $0^+_3$&$(7536)$&$\rightarrow$&$1^+_1$&$(4695)$          &&\multicolumn{2}{c}{} &  0.185\!\pm\!0.018        \\
    2833.4&1.0  & 2832.4&1.5           && $2^+_5$&$(7115)$&$\rightarrow$&$2^+_2$&$(4282)$          &&\multicolumn{2}{c}{} &  0.019\!\pm\!0.013        \\ 
    2719.5&0.5  & 2719.0&0.5           && $1^+_2$&$(7001)$&$\rightarrow$&$2^+_2$&$(4282)$          &&\multicolumn{2}{c}{} &  0.533^{+0.019}_{-0.024}  \\
    2494.7&1.6  & 2495.2&2.3           && $1^+_3$&$(7190)$&$\rightarrow$&$1^+_1$&$(4695)$          &&\multicolumn{2}{c}{} &  0.016\!\pm\!0.014        \\ 
    2464.6&0.4  & 2464.4&0.5           && $1^+_1$&$(4695)$&$\rightarrow$&$2^+_1$&$(2230)$          && 4.0&0.4             &  4.24\!\pm\!0.05          \\
    2419.9&1.1  & 2417.7&0.6           && $2^+_5$&$(7115)$&$\rightarrow$&$1^+_1$&$(4695)$          &&\multicolumn{2}{c}{} &  0.057^{+0.013}_{-0.015}  \\
    2384.2&1.0  & 2383.3&0.5           && $2^+_4$&$(6666)$&$\rightarrow$&$2^+_2$&$(4282)$          &&\multicolumn{2}{c}{} &  0.077^{+0.019}_{-0.021}  \\
    2306.0&0.6  & 2305.2&0.5           && $1^+_2$&$(7001)$&$\rightarrow$&$1^+_1$&$(4695)$          &&\multicolumn{2}{c}{} &  0.137\!\pm\!0.023        \\
   2230.49&0.15 & 2230.2&0.4           && $2^+_1$&$(2230)$&$\rightarrow$&$0^+_1$&$(\mathrm{g.s.})$ && 92&4              & 91.9^{+0.6}_{-0.4}          \\
    2051.2&0.3  & 2050.7&0.4           && $2^+_2$&$(4282)$&$\rightarrow$&$2^+_1$&$(2230)$          &&\multicolumn{2}{c}{} &  0.47\!\pm\!0.04          \\
    1970.7&1.1  & 1969.3&0.6           && $2^+_4$&$(6666)$&$\rightarrow$&$1^+_1$&$(4695)$          &&\multicolumn{2}{c}{} &  0.15\!\pm\!0.04          \\
    1770.0&1.4  & 1769.6&0.4           && $2^+_3$&$(5549)$&$\rightarrow$&$0^+_2$&$(3778)$          &&\multicolumn{2}{c}{} &  0.136\!\pm\!0.026        \\
    1641.6&1.8  & \multicolumn{2}{c}{} && $1^+_3$&$(7190)$&$\rightarrow$&$2^+_3$&$(5549)$          &&\multicolumn{2}{c}{} & <0.04                     \\ 
    1566.8&1.4  & \multicolumn{2}{c}{} && $2^+_5$&$(7115)$&$\rightarrow$&$2^+_3$&$(5549)$          &&\multicolumn{2}{c}{} & <0.030                    \\ 
    1547.8&1.0  & 1547.1&0.4           && $0^+_2$&$(3778)$&$\rightarrow$&$2^+_1$&$(2230)$          && 3.6&0.6             &  3.155^{+0.040}_{-0.036}  \\
    1452.9&1.1  & 1451.8&0.4           && $1^+_2$&$(7001)$&$\rightarrow$&$2^+_3$&$(5549)$          &&\multicolumn{2}{c}{} &  0.276\!\pm\!0.019        \\
    1266.7&1.0  & 1265.7&0.6           && $2^+_3$&$(5549)$&$\rightarrow$&$2^+_2$&$(4282)$          &&\multicolumn{2}{c}{} & <0.036\!\pm\!0.013        \\ 
    1117.6&1.4  & \multicolumn{2}{c}{} && $2^+_4$&$(6666)$&$\rightarrow$&$2^+_3$&$(5549)$          &&\multicolumn{2}{c}{} & <0.022                    \\ 
     916.9&1.1  &  915.8&0.5           && $1^+_1$&$(4695)$&$\rightarrow$&$0^+_2$&$(3778)$          &&\multicolumn{2}{c}{} &  0.034\!\pm\!0.009        \\
     853.2&1.1  &  851.8&0.5           && $2^+_3$&$(5549)$&$\rightarrow$&$1^+_1$&$(4695)$          &&\multicolumn{2}{c}{} &  0.027\!\pm\!0.008        \\
     503.4&1.0  & \multicolumn{2}{c}{} && $2^+_2$&$(4282)$&$\rightarrow$&$0^+_2$&$(3778)$          &&\multicolumn{2}{c}{} & <0.04                     \\ 
     413.5&0.5  & \multicolumn{2}{c}{} && $1^+_1$&$(4695)$&$\rightarrow$&$2^+_2$&$(4282)$          &&\multicolumn{2}{c}{} & <0.019                    \\ 
    \hline
    \hline
    \multicolumn{14}{l}{\ }\\[-0.75em]
    \multicolumn{14}{l}{$^a$Calculated from the adopted levels of Ref.~\cite{NuclDataSheets}.}\\
    \multicolumn{14}{l}{$^b$Ref.~\cite{detraz}.}    
  \end{tabular}
\end{table*}

There is generally good agreement with the results of 
D\'etraz \emph{et al.}, although we find significantly less strength in the 
$\beta$ branches to the 3.78 and 4.28~MeV levels.  We attribute this 
discrepancy to the fact that many of the higher levels not considered in 
Ref.~\cite{detraz} $\gamma$ de-excite through these levels; thus though our 
$\gamma$-ray yields for these states are in good agreement, our $\gamma$-ray 
feeding from higher levels results in a smaller deduced $\beta$ branch for 
these states.  Another difference from D\'etraz \emph{et al.} is seen with 
the 7190~keV $1^+_3$ level where we see less than half as much $\gamma$ yield, 
and find a $\beta$ branch that is 30\% smaller.  It is difficult to comment on 
this discrepancy since an efficiency curve is not provided in 
Ref.~\cite{detraz}.

The shell-model predictions of the $\beta$ branches for the five highest energy 
levels of Table~\ref{table:beta-branches} are:  0.22(4)\% (7536~keV); 0.05(4)\% 
(7637~keV); 0.10(3)\% (7921~keV); 0.06(1)\% (8125~keV); and 0.06(1)\% 
(9208~keV).  The agreement is quite reasonable, where the only significant 
difference ($>2\sigma$) seen is in the branch to the 8125-keV level; here the 
shell model calculation predicts $\approx3\times$ more strength than our 
limits allow for the branch to the 8125-keV state.  Note that the sum of these 
branches in the shell model, $0.49(7)\%$, is in perfect agreement with the 
corresponding sum of the observed branches: $(0.51^{+0.10}_{-0.14})\%$.  Given 
the high density of states in this energy range, it is a testament to the 
quality of the shell-model in this case that these branches are reproduced so 
well.  It further justifies our use of the shell model to account for the 
Pandemonium effect as discussed earlier.

In addition to reducing the uncertainties in all of the branches/yields by 
approximately an order of magnitude, we have observed $22$ more $\gamma$ 
branches and three new $\beta$ branches compared to D\'etraz 
\emph{et al.}~\cite{detraz}.  For the ten $\gamma$  and two $\beta$ transitions 
where we did not observe a statistically significant branch, a 90\% CL limit 
is quoted.

\subsection{Comparison to Shell-Model Calculations}
Shell-model calculations have been performed for the states involved in the 
$\beta$-decay of \tsups{32}Cl, which has a spin-parity of $1^+$.   For 
transitions to $0^+$ and $2^+$ states in \tsups{32}S, the strength is pure 
Gamow-Teller whereas for the isobaric analogue transition, the decay is almost 
pure Fermi.  Decays to non-analogue $1^+$ states can also include a Fermi 
component (via isospin mixing) and so for an experimental branch to one of 
these states, we can proceed in one of two ways:
\begin{enumerate}
\item make some assumptions about the Fermi contribution and deduce 
  $|M(\mathrm{GT})|$.
\item make some assumptions about the Gamow-Teller contribution and deduce 
  $|M(\mathrm{F})|$.
\end{enumerate}
 
In the next section we discuss Gamow-Teller matrix 
elements after making some minimal assumptions about the Fermi contribution. 
We will compare our experimental $M(\mathrm{GT})$ values with shell-model 
computations to say something about the quality of the \usd{} wave functions.  
Following this, we take the other approach 
by assuming that the \usd{} values for $B(\mathrm{GT})$ are correct allowing us 
to deduce $B(\mathrm{F})$.  In turn we can calculate an experimental value for 
the isospin-mixing parameter, $\delta_{C1}$, which we compare with theoretical 
calculations.

\subsubsection{Gamow-Teller matrix elements}\label{sec:assumeBF}
Here we present a comparison of calculated versus experimental
$B(\mathrm{GT})$'s. Because several of the transitions are quite retarded, 
with $\log{ft}$ values exceeding 5 (see Table~\ref{table:beta-branches}), the 
spectrum shape may depart significantly from the allowed shape.  To proceed, 
we use a shell-model calculation to compute the shape correction function 
$C(W)$ as described in the appendix of Ref.~\cite{Hardy:05}.  We define an 
``exact'' statistical rate function as
\begin{align}
  f_\mathrm{exact} = \int_1^{W_0} p W (W_0 - W)^2\, F(Z,W)\, C(W)\, dW ,
  \label{eq:fexact}
\end{align}
where $W=E_e/m_e$ is the electron total energy in electron rest-mass units, 
$W_0$ is the maximum value of $W$, $p=(W^2-1)^{1/2}$ is the electron momentum, 
$Z$ is the charge of the daughter nucleus, and $F(Z,W)$ is the Fermi 
function.  The usual statistical rate function, $f$, as used for example in 
Table~\ref{table:beta-branches} to obtain $\log{ft}$ values, puts the shape 
correction function $C(W)$ to unity.  Taking the Fermi strength to be zero for 
these non-analogue transitions, we obtain an experimental value for the GT 
strength from
\begin{align}
  B_\mathrm{exp}(\mathrm{GT}) = \frac{2\mathcal{F}t^{0^+\!\!\rightarrow 0^+}}{
    f_\mathrm{exact}t(1+\delta_R^{\prime})},
  \label{eq:BGT}
\end{align}
and an associated experimental GT matrix element, $M_\mathrm{exp}(\mathrm{GT})$,
defined by:
\begin{align}
  B_\mathrm{exp}(\mathrm{GT}) = g_{A,\mathrm{eff}}^2|M_\mathrm{exp}(\mathrm{GT})|^2.
  \label{eq:MGT}
\end{align}
Because $f_{\rm exact}$ depends on the GT matrix element via $C(W)$, we use an 
iterative procedure which is explained below.  In Eq.~\eqref{eq:BGT} we use 
$\mathcal{F}t^{0^+\!\!\rightarrow0^+}=3071.81(83)$~s from the survey of Hardy and 
Towner~\cite{Hardy:09}.  The partial half-life, $t$, is calculated from the 
\tsups{32}Cl half-life of $t_{1/2}=298(1)$~ms~\cite{armini} and the branches, 
$R$, listed in Table~\ref{table:beta-branches}, corrected for a small 
electron-capture fraction, $P_\mathrm{EC}$:
\begin{align}
  t = \frac{t_{1/2}}{R}\left(1+P_\mathrm{EC}\right).
  \label{eq:tpartial}
\end{align}
The correction $\delta_R^{\prime}$ is the transition-dependent part of the 
radiative correction and is obtained from a standard QED calculation that 
depends on $Z$ and $W$.  It is evaluated to order $\alpha$ and $Z \alpha^2$, 
with the order $Z^2\alpha^3$ terms 
estimated~\cite{Sirlin:67,Sirlin:87,JR:87,CMS:04}.  Thus 
with the quantities on the right-hand side of Eq.~\eqref{eq:BGT} determined, 
an experimental $B_\mathrm{exp}(\mathrm{GT})$ value is obtained.  This relates 
to the Gamow-Teller matrix element, $M_\mathrm{exp}(\mathrm{GT})$, as shown in 
Eq.~\eqref{eq:MGT}.  It requires knowledge of the ratio of the axial-vector 
to vector coupling constants, denoted $g_A$, for which an effective value is 
used, $g_{A,\mathrm{eff}}$ in the context of shell-model calculations in finite 
model spaces.  In the $s,d$ shell, the systematic studies of Wildenthal 
and Brown~\cite{BW:85,BW:87} have shown the effective value to be of order 
unity, so we take $g_{A,\mathrm{eff}} = 1$.  

In the shell model, the calculation of the Gamow-Teller matrix element is 
based on:
\begin{align}
  M(\mathrm{GT}) = \sum_{\alpha ,\beta} \langle f | a_{\alpha}^{\dag} b_{\beta} |
  i \rangle \langle \alpha | \mathrm{GT} | \beta \rangle ,
  \label{eq:MGTsm}
\end{align}
where $a_{\alpha}^{\dag}$ creates a neutron in quantum state $\alpha$, $b_{\beta}$ 
annihilates a proton in quantum state $\beta$ and $\langle\alpha|\mathrm{GT}|
\beta\rangle$ is the single-particle Gamow-Teller matrix element.  The matrix 
element of $ a_{\alpha}^{\dag} b_{\beta} $ in the initial and final many-body 
states are known as the one-body density matrix elements (OBDME).  The same 
OBDME used in the construction of the shape-correction function, $C(W)$, are 
also used in the shell-model evaluation of $M(\mathrm{GT})$.  Our procedure is 
to scale one of these OBDME, recompute $C(W)$ and $f_\mathrm{exact}$, and 
obtain a new $M_\mathrm{exp}(\mathrm{GT})$ from Eq.~\eqref{eq:BGT}.  We then 
repeat the procedure, refining the scaling at each step until the theory 
input matches the experimental output.  We found that it did not matter which 
\usd{} calculation we started from, or which OBDME we scaled; the convergent 
result for the Gamow-Teller matrix element was always the same.  Thus the 
method is quite stable.  

\begin{table*}
  \caption{Deduced experimental Gamow-Teller matrix elements, 
    $M_\mathrm{exp}(\mathrm{GT})$, in the decay of \tsups{32}Cl, and comparison 
    with three shell-model calculations. The RMS deviations in the predicted 
    $|M(\mathrm{GT})|$ compared to experiment are $0.073$ (\usd), $0.068$ 
    (\usda) and $0.067$ (\usdb).\label{table:MGT}}
  \begin{tabular}{D{.}{.}{6.2}@{$\pm$}D{.}{.}{4}
      c@{\hspace*{1em}}
      r@{}l@{\hspace*{1em}}
      r@{}l@{\hspace*{1em}}
      D{.}{.}{3.6}
      D{.}{.}{4.3}@{}D{.}{.}{5}
      D{.}{.}{3.5}
      D{.}{.}{3.5}
      D{.}{.}{3.5}}
    \hline
    \hline
    \multicolumn{13}{c}{  }\\[-0.8em]
    \multicolumn{8}{c}{}&
    \multicolumn{2}{c}{Experiment} &
    \multicolumn{3}{c}{Theory $|M(\mathrm{GT})|$} \\
    \multicolumn{2}{c}{$E_x$ in \tsups{32}S$^a$} & $J_n^{\pi}, T$ &
    \multicolumn{2}{c}{$t$~[s]\ \ } &
    \multicolumn{2}{c}{$f_\mathrm{exact}\,^b$} &
    \multicolumn{1}{c}{$\delta_R^{\prime}~[\%]$} &
    \multicolumn{2}{c}{$|M(\mathrm{GT})|$} &
    \multicolumn{1}{c}{\usd} &
    \multicolumn{1}{c}{\usda} &
    \multicolumn{1}{c}{\usdb} \\
    \hline
    \multicolumn{13}{c}{  }\\[-0.8em]
   9207.1&0.7   & $1_5^+,1$ &\ \ 135&$_{-22}^{+115}$ &    133&$.2\!\pm\!0.2$ & 1.70(3)&  0.58&\multicolumn{1}{l}{$\!\!\!\!^{+0.05}_{-0.15}$}  & 0.300 & 0.348 & 0.298 \\
   8125.4&0.2   & $1_4^+,1$ &\ \ $>150$&          &    689&$.2\!\pm\!2.7$ & 1.55(3)& <0.20&              & 0.123 & 0.146 & 0.134 \\
   7921.8&0.9     & $0_5^+,0$ &\ \ 900&$_{-300}^{+600}$&  879&$.5\!\pm\!1.3$ & 1.52(3)& 0.087&\pm0.019      & 0.130 & 0.148 & 0.183 \\
   7637.0&1.0   & $0_4^+,0$ &\ \ $>1200$&         &   1261&$\pm4$         & 1.49(3)& <0.05&              & 0.121 & 0.022 & 0.056 \\
   7535.6&0.9   & $0_3^+,1$ &\ \ 161&$\pm18$      &   1392&$\pm3$         & 1.48(3)& 0.164&\pm0.009      & 0.193 & 0.165 & 0.192 \\
   7190.3&1.1   & $1_3^+,0$ &\ \ 48&$\pm4$        &   2024&$\pm4$         & 1.44(3)& 0.250&\pm0.010^c    & 0.143 & 0.117 & 0.162 \\
   7115.1&0.8   & $2_5^+,1$ &    48&$\pm3$        &   2232&$.4\!\pm\!2.4$ & 1.44(3)& 0.238&\pm0.008      & 0.259 & 0.166 & 0.228 \\
   7001.4&0.4   & $1_2^+,1$ &     1&.327$\pm0.012$&   2413&$.0\!\pm\!1.7$ & 1.42(3)&\multicolumn{2}{c}{} & 0.012 & 0.064 & 0.036 \\
   6665.9&0.8   & $2_4^+,0$ &    14&.3$\pm0.5$    &   3411&$\pm4$         & 1.40(3)& 0.353&\pm0.006      & 0.428 & 0.273 & 0.365 \\
   5548.4&0.8   & $2_3^+,0$ &     7&.79$\pm0.23$  &   8583&$\pm7$         & 1.30(3)& 0.301&\pm0.004      & 0.227 & 0.333 & 0.302 \\
   4695.3&0.4   & $1_1^+,0$ &     4&.89$\pm0.07$  &  15702&$\pm30$        & 1.24(3)& 0.281&\pm0.002^c    & 0.280 & 0.309 & 0.346 \\
   4281.81&0.28 & $2_2^+,0$ &    13&.7$\pm0.5$    &  21280&$\pm10$        & 1.22(3)& 0.145&\pm0.003      & 0.071 & 0.059 & 0.085 \\
   3778.3&0.7   & $0_2^+,0$ &    31&$\pm3$        &  28009&$\pm26$        & 1.18(3)& 0.083&\pm0.003      & 0.076 & 0.137 & 0.077 \\
   2230.56&0.14 & $2_1^+,0$ &     0&.504$\pm0.003$&  67470&$\pm40$        & 1.11(3)& 0.423&\pm0.001      & 0.423 & 0.421 & 0.457 \\
  \multicolumn{2}{c}{ground state} & $0_1^+,0$ &    30&$_{-5}^{+30}$   & 167900&$\pm1600$
                                                 & 0.94(3)& 0.035&\multicolumn{1}{l}{$\!_{-0.010}^{+0.003}$} & 0.004 & 0.089 & 0.024 \\[1mm]
    \hline
    \hline
    \multicolumn{13}{c}{  }\\[-0.8em]
    \multicolumn{13}{p{0.95\textwidth}}{$^a$The average of currently accepted 
      values and the present work, \emph{i.e.}\ the third column of 
      Table~\ref{table:beta-branches}.}\\
    \multicolumn{13}{p{0.95\textwidth}}{$^b$The error bar reflects both the 
      uncertainty in the $Q$-value and the range obtained for different 
      shell-model calculations of the shape-correction function, $C(W)$.}\\
    \multicolumn{13}{p{0.95\textwidth}}{$^c$For these non-analogue $1^+$ 
      transitions the experimental derived value is $\sqrt{B(\mathrm{F})+
        B(\mathrm{GT})}$, where the Fermi contribution originates from isospin 
      mixing as discussed in the text.  For the 4695-keV level, the Fermi 
      contribution is theoretically expected to be very small, so the value 
      quoted is likely a good estimate for $|M_\mathrm{exp}(\mathrm{GT})|$.  
      For the 7190-keV level, the Fermi contribution is likely substantial, 
      so a conservative estimate for $|M_\mathrm{exp}(\mathrm{GT})|$ is in the 
      range $0 < |M_\mathrm{exp}(\mathrm{GT})| < 0.260$.}
  \end{tabular}
\end{table*}

\begin{figure}\centering
  \includegraphics[angle=90,width=0.485\textwidth]{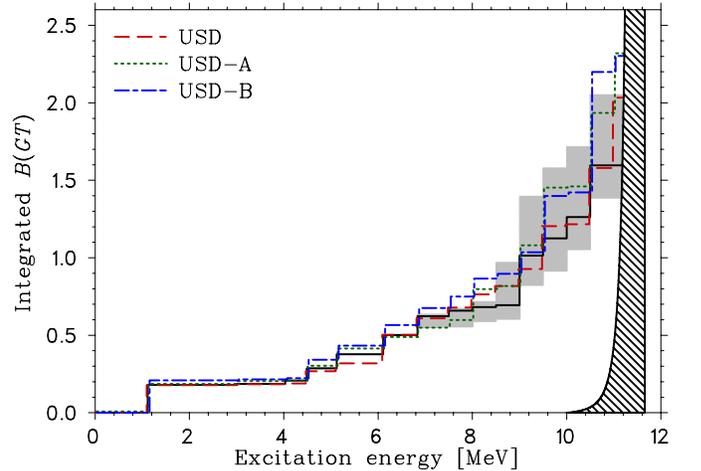}
  \caption{(Colour online) The observed integrated $B(\mathrm{GT})$ as a 
    function of excitation energy for the decay of \tsups{32}Cl and 
    comparison to shell-model calculations.  For the experimental histogram 
    (thick solid line with the shaded region representing its uncertainty 
    bars), we have assumed zero Fermi strength in the 4695-keV state, zero 
    Gamow-Teller strength in the IAS state, the range of Gamow-Teller strength 
    given in the footnote of Table~\ref{table:MGT} for the 7190-keV state, and 
    included the $\beta$-delayed particle emitting states from 
    Ref.~\cite{honkanenNPA}.  The hatched region indicates strength that 
    may be missed in the experiment because the phase space for $\beta$ 
    decay is too small resulting in $\beta$-delayed branches less than 
    $0.0015\%$ which would not have been reported in Ref.~\cite{honkanenNPA}.
    \label{fig:integrated-B(GT)}}
\end{figure}

In Table~\ref{table:MGT} we list the partial half-lives, $t$, the statistical 
rate function $f_\mathrm{exact}$ at convergence, the radiative correction 
$\delta_R^{\prime}$, and the deduced experimental Gamow-Teller matrix element.  
We also list the theoretical values from three shell-model calculations with 
effective interactions \usd, \usda{} and \usdb, now without any of the 
adjustments to the OBDME discussed above.  Thus, Table~\ref{table:MGT} 
compares $M_\mathrm{exp}(\mathrm{GT})$ values with theoretical expectations.  
The comparison is very favourable with the RMS difference $\approx0.07$.  
Where the matrix element is quite large, say $M(\mathrm{GT})>0.2$, theory 
does exceedingly well.  For retarded transitions with $M(\mathrm{GT})<0.2$, 
theory does not perform as well, but here the small values are a consequence 
of cancellations among shell-model amplitudes which are much harder to get 
precisely right.  The same comparison is presented in 
Fig.~\ref{fig:integrated-B(GT)}, where the integrated 
$B(\mathrm{GT})$ values are displayed as a function of excitation energy.  As 
has been found by Brown and Wildenthal~\cite{BW:85,BW:87}, the \usd{} 
effective interaction in $s,d$-shell nuclei gives a reasonably accurate 
picture of Gamow-Teller properties in these nuclei.  The newer interactions, 
\usda{} and \usdb, perform equally well. 

\subsubsection{Isospin-symmetry breaking in Fermi transitions}
\label{sec:assumeBGT}
We switch our attention to the $\beta$ transition to the 7001-keV, 
$1^+_2,T\!=\!1$ isobaric analogue state (IAS).  This transition is a 
mix of Fermi and Gamow-Teller components; therefore from the partial half-life 
alone it is not possible to deduce the Gamow-Teller matrix element.  Hence the 
gap in Table~\ref{table:MGT}.  However, the shell-model calculation for the 
Gamow-Teller matrix element predicts for the IAS state a very small value 
indeed.  This is a fortunate happenstance: it gives us the opportunity to 
study this transition as if it were a pure Fermi type, compare it with the 
precisely measured pure Fermi transitions between $0^+$ states, and deduce 
the amount of isospin-symmetry breaking (ISB) in this transition.  A fairly 
large ISB effect is anticipated because in \tsups{32}S the IAS state is 
only 189~keV away from the 7190-keV state; a state with the same spin but 
different isospin.  Perturbation theory predicts that when two states of the 
same spin are close together in the spectrum, Coulomb and charge-dependent 
nuclear forces induce a degree of isospin-symmetry breaking that is 
inversely proportional to the square of the energy separation of the two 
states.  For a separation of 189~keV, mixing at the several percent 
level can be anticipated.

The partial half-life for decay to the IAS state has been determined with $1\%$ 
accuracy, namely $t^\mathrm{IAS} = 1.331(12)$~s.  From this, the ISB correction 
$\delta_C$ can be determined to $15\%$ accuracy from the equation:
\begin{align}
  \begin{split}
    f_\mathrm{exact}\,t^\mathrm{IAS} & (1 + \delta_R^{\prime})
    (1 + \delta_\mathrm{NS} - \delta_C) \\
    & = \frac{K}{G_V^2 (1 + \Delta_R^V) [B(\mathrm{F}) + B(\mathrm{GT})]}\\
    & = \frac{2 \mathcal{F}t^{0^+\!\!\rightarrow 0^+}}{B(\mathrm{F})+B(\mathrm{GT})}.
  \end{split}
  \label{eq:ftIAS}
\end{align}
Here $K/(\hbar c)^6=2\pi^3\hbar\ln{2}/(m_ec^2)^5$ is a constant and $G_V$ is the 
vector coupling constant characterizing the strength of the vector weak 
interaction.  The quantity $K/G_V^2(1+\Delta_R^V)$ is taken from the 
precision work on \zerotozero{} superallowed transitions and is 
expressed in terms of $\mathcal{F}t^{0^+\!\!\rightarrow 0^+}$ introduced in 
Eq.~\eqref{eq:BGT}.  The radiative correction has been split into three 
pieces: (a) a nucleus-independent term, $\Delta_R^V$, is included in 
$\mathcal{F}t^{0^+\!\!\rightarrow 0^+}\!$; (b) a trivially nucleus-dependent 
term, $\delta_R^{\prime}$, is calculated to be $1.421(32)\%$; and (c) a second 
nucleus-dependent term, $\delta_\mathrm{NS}$, is small but requires a 
nuclear-structure calculation to be evaluated.  It is convenient to place 
$\delta_\mathrm{NS}$ and $\delta_C$ together as both are dependent on 
shell-model nuclear-structure calculations.  Finally $B(\mathrm{F})$ is the 
square of the Fermi matrix element, $B(\mathrm{F})=|M_0|^2=2$ for $T\!=\!1$ 
transitions in the isospin-symmetry limit, and $B(\mathrm{GT})$ is the square 
of the Gamow-Teller matrix element, Eq.~\eqref{eq:BGT}.  For $B(\mathrm{GT})$, 
we take the three theoretical values from the shell-model calculation using 
\usd, \usda{} 
and \usdb{} effective interactions, average them and assign an uncertainty 
equal to half the spread between the largest and smallest calculated values: 
$B(\mathrm{GT})=0.002\pm0.002$.  On rearranging Eq.~\eqref{eq:ftIAS}, we obtain
\begin{align}
  (\delta_C-\delta_\mathrm{NS})_\mathrm{exp} & = 1-\frac{2 
    \langle\mathcal{F}t^{0^+\!\!\rightarrow 0^+}\rangle}{
    f_\mathrm{exact}\,t^\mathrm{IAS} 
    (1+\delta_R^{\prime})\big[B(\mathrm{F})+B(\mathrm{GT})\big]}\nonumber \\
  & = 5.4(9)\% ,\label{eq:dCexpt}
\end{align}
a substantial isospin symmetry breaking term, the largest yet determined in a 
superallowed Fermi transition.

In what follows we present a shell-model calculation of $\delta_C-
\delta_\mathrm{NS}$ following the procedures developed by Towner and 
Hardy~\cite{TownerHardy:08}.  First, for the ISB correction $\delta_C$
defined in Eq.~\eqref{eq:defn_of_delta_C}, the technique is to introduce 
Coulomb and 
other charge-dependent terms into the shell-model Hamiltonian.  However, 
because the Coulomb force is long range, the shell-model space has to be very 
large indeed to include all the potential states with which the Coulomb 
interaction might connect.  Currently this is not a practical proposition.  
To proceed, Towner and Hardy~\cite{TownerHardy:08} divide $\delta_C$ into two 
parts:
\begin{align}
  \delta_C = \delta_{C1} + \delta_{C2}.\label{eq:dc1c2}
\end{align}
For $\delta_{C1}$, we perform a shell-model calculation in the truncated
$0\hbar\omega$ model space of the $s,d$-shell orbitals.  Charge-dependent 
terms are added to the charge-independent Hamiltonians of \usd, \usda{} and 
\usdb.  The strengths of these charge-dependent terms are adjusted to reproduce 
the $b=-5.4872(35)$~MeV and $c=0.1953(37)$~MeV~\cite{britz} coefficients of 
the isobaric multiplet mass equation (IMME) as applied to the $1^+,T\!=\!1$ 
states in $A=32$, the triplet of states involved in the $\beta$-transition 
under study.  As mentioned already, the bulk of the isospin mixing in the IAS 
occurs with the neighbouring $1^+_3$ state.  This observation is used to 
constrain and refine the calculation.  In the limit of two-state mixing, 
perturbation theory indicates that
\begin{align}
  \delta_{C1} \propto 1 / (\Delta E)^2,\label{eq:dC1dE}
\end{align}
where $\Delta E$ is the energy separation of the analogue and non-analogue 
$1^+$ states.  Thus it is important that the shell-model Hamiltonian produce a 
good-quality spectrum of $1^+$ states.  The shell model calculation has 
varying degrees of success in this regard.  For the $1^+$ states in $A=32$, 
the separation between the IAS and the third $1^+_3,T\!=\!0$ state is observed 
to be $188.9\pm1.2$~keV.\ \ 
The shell model calculates this separation to be 184~keV with 
\usd, 248~keV with \usda{} and 387~keV with \usdb{} interactions.  These are 
quite respectable results given the inherent accuracy of a shell-model 
calculation for predicting energies.  However, for a reliable $\delta_{C1}$ 
calculation, this spread in $\Delta E$ values is quite a problem.  To cope 
with this, the Towner-Hardy 
recommended procedure is to scale the calculated $\delta_{C1}$ value by a 
factor of $(\Delta E)_\mathrm{theo}^2 / (\Delta E)_\mathrm{exp}^2$, the ratio of 
the square of the energy separation of the $1^+$ states in the model 
calculation to that known experimentally.  After this is done, the 
$\delta_{C1}$ values obtained in the three shell-model calculations are 
reasonably consistent:  $\delta_{C1}=3.73\%$ for \usd, $3.32\%$ for \usda, and 
$4.19\%$ for \usdb.  We average these three results and assign an uncertainty 
equal to half the spread between them to arrive at:
\begin{align}
  \delta_{C1} = 3.75(45)\%.\label{eq:dC1valu}
\end{align}

For the calculation of $\delta_{C2}$ we need to consider mixing with states 
outside the $0\hbar\omega$ shell-model space.  The principal mixing is with 
states that have one more radial node.  Such mixing effectively changes the 
radial function of the proton involved in the $\beta$ decay relative to that of 
the neutron.  The practical calculation, therefore, involves computing radial 
overlap integrals with modeled proton and neutron radial functions.  Details 
of how this is done are given in Ref.~\cite{TownerHardy:08}.  The radial 
functions are taken to be eigenfunctions of a Saxon-Woods potential whose 
strength is adjusted so that the asymptotic form of the radial function has 
the correct dependence on the separation energy.  The initial and final 
$A$-body states are expanded in a complete set of $(A\!-\!1)$-parent states.  
The separation energies are the energy differences between the $A$-body state 
and the $(A\!-\!1)$-body parent states.  A shell-model calculation is required 
to give the spectrum of parent states and the spectroscopic amplitudes of the 
expansion.  For the three \usd{} interactions, we compute $\delta_{C2}=0.827\%$ 
for \usd{} and $0.865\%$ for both \usda{} and \usdb.  Our adopted value is:
\begin{align}
  \delta_{C2} = 0.85(3)\%.\label{eq:dC2valu}
\end{align}
The uncertainty, calculated in the same manner as described in 
Ref.~\cite{TownerHardy:08}, represents the range of results for the \usd{} 
interactions, the different methodologies considered in adjusting the strength 
of the Saxon-Woods potential, and the uncertainty in the Saxon-Woods radius 
parameter which was fitted to the experimental charge radius of \tsups{32}S.

Finally, we need an evaluation of the nuclear-structure-dependent piece of the
radiative correction, $\delta_\mathrm{NS}$.  Such a term arises because in a 
many-body system such as a nucleus, the electromagnetic interaction and the 
weak interaction that collectively induce a radiative correction do not have 
to interact with the same nucleon in the nucleus.  When these interactions 
occur with different nucleons, the process is described by two-body operators. 
The evaluation of matrix elements of two-body operators depends in detail on 
the nuclear structure of the states involved.  Such calculations were first 
made in 1992~\cite{Towner:92,BBJR:92} and updated two years 
later~\cite{Towner:94}.  We follow the latter reference and compute 
$\delta_\mathrm{NS}$ for each of the $s,d$-shell effective \usd{} interactions.  
Essentially the same result was obtained in each case.  We adopt the average 
value of
\begin{align}
  \delta_\mathrm{NS} = -0.15(2)\%.\label{eq:dNSvalu}
\end{align}
The result is a very small correction, about 3 times smaller than the 
uncertainty in $\delta_C$.  

\begin{figure}\centering
  \includegraphics[angle=90,width=0.485\textwidth]{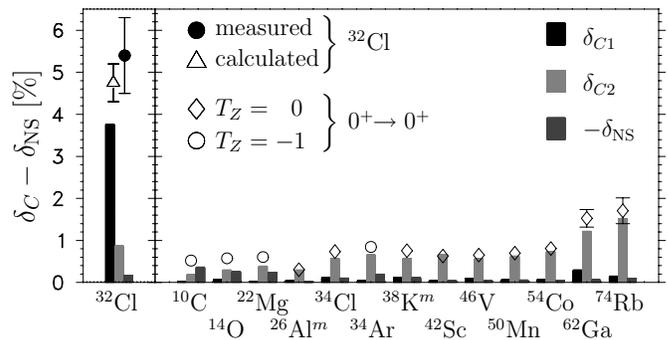}
  \caption{Our determination of the isospin-symmetry-breaking correction for 
  \tsups{32}Cl (filled circle) and calculations for \tsups{32}Cl as well as 
  other superallowed transitions (open points).  The three components, 
  $\delta_{C1}$, $\delta_{C2}$ and $\delta_\mathrm{NS}$, are shown separately.  
  The measurement and prediction for \tsups{32}Cl, particularly the 
  $\delta_{C1}$ component, is significantly larger than in any of the 
  \zerotozero{} transitions.\label{fig:deltaC-deltaNS}}
\end{figure}

Adding together Eqs.~\eqref{eq:dC1valu}, \eqref{eq:dC2valu} and 
\eqref{eq:dNSvalu}, we obtain
\begin{align}
  \delta_C - \delta_\mathrm{NS} = 4.8(5)\%,\label{eq:dCtheo}
\end{align}
which agrees with the experimental result of $5.4(9)\%$ of
Eq.~\eqref{eq:dCexpt} within stated uncertainties.  A comparison of this result 
with calculations for the \zerotozero{} cases is shown in 
Fig.~\ref{fig:deltaC-deltaNS}.  As one can clearly see, the correction in 
\tsups{32}Cl is about five times larger than the typical $<1\%$ values found 
for the $s,d$-shell nuclei in \zerotozero{} superallowed transitions.  
The TH model, which has already been shown~\cite{TH-PRC82-2010} to reproduce 
the nucleus-to-nucleus variation of ISB effects in superallowed zerotozero{} 
transitions required by the CVC hypothesis, is shown here to produce a 
much larger ISB effect as again verified by the current experiment.

Let us now briefly consider the isospin mixing in the non-analogue $1^+$ states 
at $7190$ and $4695$~keV, and deduce experimental values for the degree of 
mixing in the same way as just described for the IAS state.  We assume the 
shell-model calculation for $B(\mathrm{GT})$ to be correct, assigning an 
uncertainty equal to half the spread between the different results obtained 
with the  \usd, \usda{} and \usdb{} interactions.  For the $7190$-keV state, 
the shell-model calculations yield $B(\mathrm{GT})=0.020(6)$, and for the 
$4695$-keV state $B(\mathrm{GT})=0.098(21)$.  An experimental value for 
$B(\mathrm{F})+B(\mathrm{GT})$ is computed from a rearranged 
Eq.~\eqref{eq:dCexpt}:
\begin{align}
  B_\mathrm{exp}(\mathrm{F}) + B_\mathrm{exp}(\mathrm{GT}) = 
  \frac{2\mathcal{F}t^{0^+\!\!\rightarrow 0^+}}
  {f_\mathrm{exact} t (1 + \delta_R^{\prime})}, \label{eq:BFnon}
\end{align}
where the nuclear-structure-dependent radiative correction, $\delta_\mathrm{NS}$,
is ignored.  Inserting the experimental values into the right-hand side of 
Eq.~\eqref{eq:BFnon}, we obtain 
\begin{align}
  B_\mathrm{exp}(\mathrm{F}) + B_\mathrm{exp}(\mathrm{GT}) & = 0.062\pm0.005 
  \nonumber \\
  B_\mathrm{exp}(\mathrm{F}) & = 0.042\pm0.008\label{eq:BF9}
  \intertext{for the 7190-keV state, and}
  B_\mathrm{exp}(\mathrm{F}) + B_\mathrm{exp}(\mathrm{GT}) & = 0.0790\pm0.0012 
  \nonumber \\
  B_\mathrm{exp}(\mathrm{F}) & = -0.019\pm0.021 \nonumber \\
  \mathrm{or}\qquad B_\mathrm{exp}(\mathrm{F}) & <0.014
  \label{eq:BF4}
\end{align}
for the 4695-keV state.  In the limit of exact isospin symmetry the 
$B(\mathrm{F})$ values would be zero for these non-analogue transitions.  So 
the non-zero value in Eq.~\eqref{eq:BF9} is a further indication that 
isospin-symmetry breaking is present.  For non-analogue transitions, we define
\begin{align}
  B(\mathrm{F}) = |M_0|^2 \delta_{C1}^n (1 - \delta_{C2}) \simeq 2 \delta_{C1}^n 
  \label{eq:BFmix}
\end{align}
where $\delta_{C1}^n$ is the isospin-symmetry breaking correction for the 
$n$\tsups{th} $1^+$ state in \tsups{32}S computed in the $s,d$ shell-model
space, and $\delta_{C2}$ the radial overlap correction representing Coulomb 
mixing beyond the $0\hbar\omega$ model space. In Eq.~\eqref{eq:BFmix}, we 
drop the $\delta_{C1}\delta_{C2}$ cross term as being negligible.  
From the experimental 
results in Eqs.~\eqref{eq:BF9} and \eqref{eq:BF4}, we determine 
$\delta_{C1,\,\mathrm{exp}}^3=2.1(4)\%$ for the 7190-keV state and 
$\delta_{C1,\,\mathrm{exp}}^1<0.7\%$ for the 4695-keV state.  The 
theory calculation that produced the result for $\delta_{C1}$ in 
Eq.~\eqref{eq:dC1valu} also gives as a by-product values of $\delta_{C1}^n$ for 
the non-analogue transitions.  For the 7190-keV and 4695-keV states, theory 
predicts $\delta_{C1}^3=3.2(3)\%$ and $\delta_{C1}^1=0.04(1)\%$ respectively.  
Evidently there is a small discrepancy between theory and experiment for the 
symmetry-breaking in the 7190-keV transition.  On the theory side, if we 
accept the symmetry-breaking in the IAS state to be correct, then it is most 
likely to be correct in the 7190-keV state as well.  This is because the 
situation is close to 2-state mixing, and the loss of Fermi strength to the 
IAS state is recovered in the 7190-keV state.  On the experimental side, the 
result depends on the correctness of the Gamow-Teller strength calculated with 
\usd{} wave functions.  If the $B(\mathrm{GT})$ were over-estimated by the 
\usd{} calculation, then theoretical and experimental values for the isospin 
symmetry mixing in the 7190-keV state could easily be reconciled. 

\section{Conclusions}
We have measured relative $\gamma$-ray intensities and deduced $\beta$-decay 
branches for the decay of \tsups{32}Cl. We have observed 3 new $\beta$ 
branches, 22 new $\gamma$ lines, placed limits on 2 other $\beta$ branches 
and 10 other $\gamma$ transitions, and have improved the precision on 
previously known yields and branches by about an order of magnitude. 

In total, twelve $\beta$ branches have been measured in the decay of 
\tsups{32}Cl.  Eleven of these are Gamow-Teller transitions and one is 
predominantly Fermi.  For the Gamow-Teller transitions, the GT matrix element 
has been determined and compares favourably with shell-model calculations 
using \usd{} effective interactions.  These calculations also find the 
Gamow-Teller component in the IAS transition to be very small, indicating this 
transition is almost pure Fermi-like.  Thus, this transition can be analyzed 
in an identical way to that used for the \zerotozero{} superallowed 
transitions.  We extract a sizable isospin symmetry breaking correction for 
this transition, $\delta_C - \delta_\mathrm{NS} = 5.4(9)\%$, which agrees well 
with a theoretical value of $4.8(5)\%$.  

In addition, the improved precision in the relative $\gamma$-ray intensities 
can be used for a more precise determination of $\gamma$-ray efficiencies in 
the decay of \tsups{32}Ar~\cite{ar32-paper}.  The $\gamma$ intensity from the 
lowest $T\!=\!2$ state in \tsups{32}Cl is of interest for measuring isospin 
symmetry breaking in the $T\!=\!2$ superallowed decay of \tsups{32}Ar.  
Presently the $\gamma$-decay intensities from the decay of \tsups{32}Ar are 
limited by statistical precision, but the present work opens the possibility 
of determining its $\gamma$ branches to higher precision in future experiments.

\section{Acknowledgments}
We acknowledge the support staff of the Cyclotron Institute and the Center 
for Experimental Nuclear Physics and Astrophysics.  The work of the Texas 
A\&M authors is supported by the U.S. Department of Energy under Grant 
No.\ DE-FG02-93ER40773 and by the Robert A. Welch Foundation under Grant 
No.\ A-1397.  The University of Washington authors were supported by 
the U.S. Department of Energy under Grant No.\ DE-FG02-97ER41020.
%
%

\end{document}